\documentclass[preprint2]{proto}
\usepackage{xspace} 
\usepackage{natbib}
\bibpunct{(}{)}{;}{a}{,}{,}
\usepackage{times}

\newcommand{\spitzer}{\textit{Spitzer}\xspace}
\newcommand{\herschel}{\textit{Herschel}\xspace}

\newcommand{\lsun}{\mbox{L$_\odot$}}
\newcommand{\msun}{\mbox{M$_\odot$}}
\newcommand{\rsun}{R$_{\odot}$}
\newcommand{\um}{$\mu$m}
\newcommand{\kms}{km s$^{-1}$}
\newcommand{\lbol}{$L_{\rm bol}$}
\newcommand{\lint}{$L_{\rm int}$}
\newcommand{\tbol}{$T_{\rm bol}$}

\newcommand{\lsl}{$L_{\rm smm}$/$L_{\rm bol}$}
\newcommand{\lsmm}{$L_{\rm smm}$}

\newcommand{\lsmmbol}{$L_{\rm smm}$/$L_{\rm bol}$}

\newcommand{\cojone}{$^{12}$CO (1--0)}
\newcommand{\cotwo}{\mbox{CO$_2$}}

\newcommand{\beqa}{\begin{eqnarray}}
\newcommand{\eeqa}{\end{eqnarray}}
\newcommand{\avg}[1]  {{\langle #1 \rangle}} 
\newcommand{\beq}	{\begin{equation}}
\newcommand{\eeq}	{\end{equation}}

\newcommand{\tf} 	{{t_f}}
\newcommand{\mf} 	{{m_f}}

\begin{document}

\title{\textbf{\LARGE The Evolution of Protostars: Insights from \\
Ten Years of Infrared Surveys with {\it Spitzer} and {\it Herschel}}}

\author {\textbf{\large Michael M.~Dunham}}
\affil{\small\em Yale University; Harvard-Smithsonian Center for Astrophysics}
\author {\textbf{\large Amelia M.~Stutz}}
\affil{\small\em Max Planck Institute for Astronomy}
\author {\textbf{\large Lori E.~Allen}}
\affil{\small\em National Optical Astronomy Observatory}
\author {\textbf{\large Neal J.~Evans II}}
\affil{\small\em The University of Texas at Austin}
\author {\textbf{\large William J.~Fischer and S.~Thomas Megeath}}
\affil{\small\em The University of Toledo}
\author {\textbf{\large Philip C.~Myers}}
\affil{\small\em Harvard-Smithsonian Center for Astrophysics}
\author {\textbf{\large Stella S.~R.~Offner}}
\affil{\small\em Yale University}
\author {\textbf{\large Charles A.~Poteet}}
\affil{\small\em Rensselaer Polytechnic Institute}
\author {\textbf{\large John J.~Tobin}}
\affil{\small\em National Radio Astronomy Observatory}
\author {\textbf{\large Eduard I.~Vorobyov}}
\affil{\small\em University of Vienna; Southern Federal University}

\begin{abstract}
\baselineskip = 11pt
\leftskip = 0.65in 
\rightskip = 0.65in
\parindent=1pc
{\small Stars form from the gravitational collapse of dense 
molecular cloud cores. In the protostellar phase, mass accretes from the core 
onto a protostar, likely through an accretion disk, and it is during this 
phase that the initial masses of stars and the initial conditions for 
planet formation are set.  
Over the past decade, new observational capabilities provided by the 
{\it Spitzer Space Telescope} and {\it Herschel Space Observatory} have 
enabled wide-field surveys of entire star-forming clouds with 
unprecedented sensitivity, resolution, and infrared wavelength coverage.
We review resulting advances in the field, 
focusing both on the observations themselves and the constraints they  
place on theoretical models of star formation and protostellar evolution.
We also emphasize open questions and outline new directions needed to further 
advance the field.
%
 \\~\\~\\~}

\end{abstract}  

\
\section{\textbf{INTRODUCTION}}


%
The formation of stars occurs in dense cores of molecular clouds,
where gravity finally overwhelms turbulence, magnetic fields, and 
thermal pressure.
This review assesses the current understanding of the early stages of this
process, focusing on the evolutionary progression from dense cores up to the
stage of a star and disk without a surrounding envelope: the protostellar phase
of evolution.

Because of the obscuration at short wavelengths by dust and the consequent
re-emission at longer wavelengths, protostars 
are best studied at infrared and radio wavelengths, using both
continuum emission from dust and spectral lines.
At the time of the last Protostars and Planets conference, only initial 
{\it Spitzer} results were reported and {\it Herschel} had not yet launched.  
For this review, we can incorporate much more complete {\it Spitzer} results 
and report on some initial {\it Herschel} results.  We focus this review on 
the progress made with these facilities toward answering several key questions 
about protostellar evolution, as outlined below.

How do surveys find protostars and distinguish them from background galaxies,
AGB stars, and more evolved star plus disk systems?  What is the resulting 
census of protostars and young stars in nearby molecular clouds?  Various 
identification methods have been employed and are compared in \S 
\ref{sec_surveys}, leading to a much firmer census from relatively 
unbiased surveys of clouds in the Gould Belt and Orion.

How do we distinguish various stages in protostellar evolution? Theoretically,
a collapsing core initially forms a first hydrostatic core, which quickly
collapses again to form a true protostar, referred to as Stage 0. As material
moves from the surrounding core to the disk and onto the star, the stellar
mass eventually exceeds the core mass, and the system becomes a Stage I
object. We discuss in \S \ref{sec_surveys} the observational correlatives 
to these Stages (called Classes) and the reliability of using observational 
signatures to distinguish between the Stages, while various candidates for the 
elusive first hydrostatic core are considered in \S \ref{sec_earliest}.


How long do each of the observational Classes last? Ideally, we would ask this
question about the Stages, but the issues addressed in \S \ref{sec_surveys} 
restrict current information about timescales to the Classes.
With the large surveys now available and with careful removal of contaminants,
we can now estimate durations of the observational Classes (\S 
\ref{sec_surveys},\ref{sec_earliest}).

What is the luminosity distribution of protostars (\S \ref{sec_surveys})?
What histories of accretion are predicted by various theoretical and 
semi-empirical models, and how do these compare to the observations
(\S \ref{sec_accretion})?  
Does accretion onto the star vary smoothly or is it more episodic?
This question is addressed in significant detail in the accompanying chapter 
by {\it Audard et al}.  Here we address it with respect to the luminosity 
distribution in 
\S \ref{sec_accretion}, and other observational constraints are discussed 
in \S \ref{sec_variability}.  While some results of monitoring are 
becoming available, most information on the protostellar accretion or 
luminosity variations comes from indirect proxies such as outflow patterns and
chemical effects with timescales appropriate to reflect the luminosity 
evolution.

How does the disk form and evolve during the protostellar phase and how
is this affected by the accretion history? This question is also addressed
from a theoretical perspective in the accompanying chapter by 
{\it Li et al.}, so we focus in \S \ref{sec_disks} on the observational 
evidence for and properties of disks in the protostellar phases.

What is the microscopic (chemical and mineralogical) evolution of the 
infalling material? Spectroscopy has provided information on the nature
of the building blocks of comets and planetesimals as they fall toward
the disk (\S \ref{sec_infalling}).

What is the effect of the star forming environment? Protostars are found
in a wide range of environments, ranging from isolated dense cores to
low density clusters of cores in a clump, to densely clustered regions like
Orion.  
These issues are addressed in \S \ref{sec_environment}.

The organization of this review chapter is motivated by first providing a 
very broad overview of the protostars that have been revealed by {\it Spitzer} 
and {\it Herschel} surveys (\S 2), followed by a general theoretical overview 
of the protostellar mass accretion process (\S 3).  The rest of the review 
then focuses on specialized topics related to protostellar evolution, including 
the earliest stages of evolution (\S 4), the evidence for episodic mass 
accretion in the protostellar stage (\S 5), the formation and early evolution 
of protostellar disks (\S 6), the evolution of infalling material (\S 7), 
and the role of environment (\S 8).

\section{\textbf{PROTOSTARS REVEALED BY INFRARED SURVEYS}}\label{sec_surveys}

Within the nearest 500 pc, low-mass young stellar objects (YSOs) can be detected by modern instruments across most of the initial mass function over a wide range of wavelengths.  
Due to the emission from surrounding dust in envelopes and/or disks, YSOs are best identified at infrared wavelengths where all but the most embedded protostars are detectable (see \S \ref{sec_earliest}).  Older pre-main-sequence stars that have lost their circumstellar dust are not revealed in the infrared but can be identified in X-ray surveys \citep[e.g.,][]{2007A&A...468..379A,2010AJ....140..266W,2013ApJ...768...99P}.  

Several large programs to detect and characterize YSOs have been carried out with the {\it Spitzer Space Telescope} \citep{2004ApJS..154....1W}, which operated at 3--160 \micron\ during its cryogenic mission. These include ``From Molecular Cores to Planet-Forming Disks'' \citep[c2d;][]{2003PASP..115..965E,2009ApJS..181..321E}, which covered seven large, nearby molecular clouds and $\sim$100 isolated dense cores, the {\it Spitzer} Gould Belt Survey \citep{2013AJ....145...94D}, which covered 11 additional nearby clouds, and {\it Spitzer} surveys of the Orion \citep{2012AJ....144..192M} and Taurus \citep{2010ApJS..186..259R} molecular clouds.  These same regions have recently been surveyed by a number of key programs with the far-infrared {\it Herschel Space Observatory} \citep{2010A&A...518L...1P}, which observed at 55--670 \micron, including the {\it Herschel} Gould Belt Survey \citep{2010A&A...518L.102A} and HOPS, the {\it Herschel} Orion Protostar Survey \citep{2013AN....334...53F,2013ApJ...763...83M,2013ApJ...767...36S}.  The wavelength coverage and resolution offered by {\it Herschel} allow a more precise determination of protostellar properties than ever before.  Many smaller {\it Spitzer} and {\it Herschel} surveys have also been performed and their results will be referred to when relevant.


The results from these large surveys differ due to a combination of sample selection, methodology, sensitivity, resolution, and genuine differences in the star formation environments.  While the last of these is of major scientific interest and is touched upon in \S \ref{sec_environment}, definitive evidence for differences in star formation caused by environmental factors awaits a rigorous combining of the samples and analysis with uniform techniques, which is beyond the scope of this review.  Here we highlight the broad similarities and differences in the surveys.

\subsection{Identification}

YSOs are generally identified in the infrared by their red colors relative to foreground and background stars.  Reliable identification is difficult since many other objects have similar colors, including star-forming galaxies, active galactic nuclei (AGN), background asymptotic giant branch (AGB) stars, and knots of shock-heated emission.  Selection criteria based on position in various color-color and color-magnitude diagrams have been developed to separate YSOs from these contaminants and are described in detail by \citet{2007ApJ...663.1149H} for the c2d and Gould Belt surveys and by \citet{2009ApJS..184...18G}, \citet{2009AJ....137.4072M,2012AJ....144..192M}, and \citet{2012AJ....144...31K} for the Orion survey.  

Once YSOs are identified, multiple methods are employed to pick out the subset that are protostars still embedded in and presumed to be accreting from surrounding envelopes.  For the c2d and Gould Belt clouds, \citet{2009ApJS..181..321E} and \citet{2013AJ....145...94D} defined as protostars those objects with at least one detection at $\lambda \geq$ 350 \um, with the rationale being that (sub)millimeter detections in typical surveys of star-forming regions trace surrounding dust envelopes but not disks due to the relatively low mass sensitivities of these surveys \citep[see][]{2013AJ....145...94D}.  On the other hand, \citet{2012AJ....144...31K} identified protostars in the c2d, Taurus, and Orion clouds using a set of color and magnitude criteria designed to pick out objects with the expected red colors of protostars, but they did not require (sub)millimeter detections.  A comparison of the two samples shows general agreement except for a large excess of faint protostars in the \citet{2012AJ....144...31K} sample.

This discrepancy in the number of faint protostars may be due to incompleteness in the \citet{2013AJ....145...94D} sample, whose requirement of a (sub)millimeter detection may have introduced a bias against the lowest-mass cores and thus the lowest-mass (and luminosity) protostars. It may also be due to unreliability of some detections by \citet{2012AJ....144...31K}, whose corrections for contamination from galaxies, edge-on disks, and highly extinguished Class II YSOs is somewhat uncertain, especially at the lowest luminosities where contamination steeply rises. Both effects likely contribute.  Resolution of this discrepancy should be possible in the near future once complete results from the {\it Herschel} and James Clerk Maxwell Telescope SCUBA-2 Legacy Gould Belt surveys are available, since together they will fully characterize the population of dense cores in all of the Gould Belt clouds with sensitivities below 0.1 \msun.  This general trade-off of completeness versus reliability remains the subject of ongoing study, with \citet{2013ApJS..205....5H} recently presenting a new method of identifying YSOs in {\it Spitzer} data that may increase the number of YSOs by $\sim$30\% without sacrificing reliability.  We anticipate continued developments in the coming years.

\subsection{Classification}

The modern picture of low-mass star formation resulted from the merger of an empirical classification scheme 
with theoretical work on the collapse of dense, rotating cores \citep[e.g.,][]{1987ARA&A..25...23S,1987ApJ...312..788A,1987BAAS...19R1092W,1993ApJ...414..676K}, and was reviewed by \citet{2007prpl.conf..117W} and \citet{2007prpl.conf..361A} at the last Protostars and Planets conference.  In this picture there are three stages of evolution after the initial starless core.  In the first, the protostellar stage, an infalling envelope feeds an accreting circumstellar disk. In the second, the envelope has dissipated, leaving a protoplanetary disk that continues to accrete onto the star.  In the third, the star has little or no circumstellar material, but it has not yet reached the main sequence.  Defining $S_\lambda$ as the flux density at wavelength $\lambda$, \citet{1987IAUS..115....1L} used the infrared spectral index,
\begin{equation}
\alpha=\frac{d \log (\lambda S_\lambda)}{d \log \lambda} \qquad ,
\end{equation}
to divide objects into three groups -- Class I, II, and III -- meant 
to correspond to these three stages.  
\citet{1994ApJ...434..614G} added a fourth class, the ``Flat-SED'' 
sources (see \S \ref{sec_timeline}), leading to the following classification 
system:  
\begin{itemize}
\item Class I: $\alpha \ge 0.3$ 
\vspace{-0.1in}
\item Flat-SED: $-0.3\le \alpha < 0.3$
\vspace{-0.1in}
\item Class II: $-1.6\le \alpha < -0.3$
\vspace{-0.1in}
\item Class III: $\alpha < -1.6$
\end{itemize}

\begin{figure}[!th]
\epsscale{0.9}
\plotone{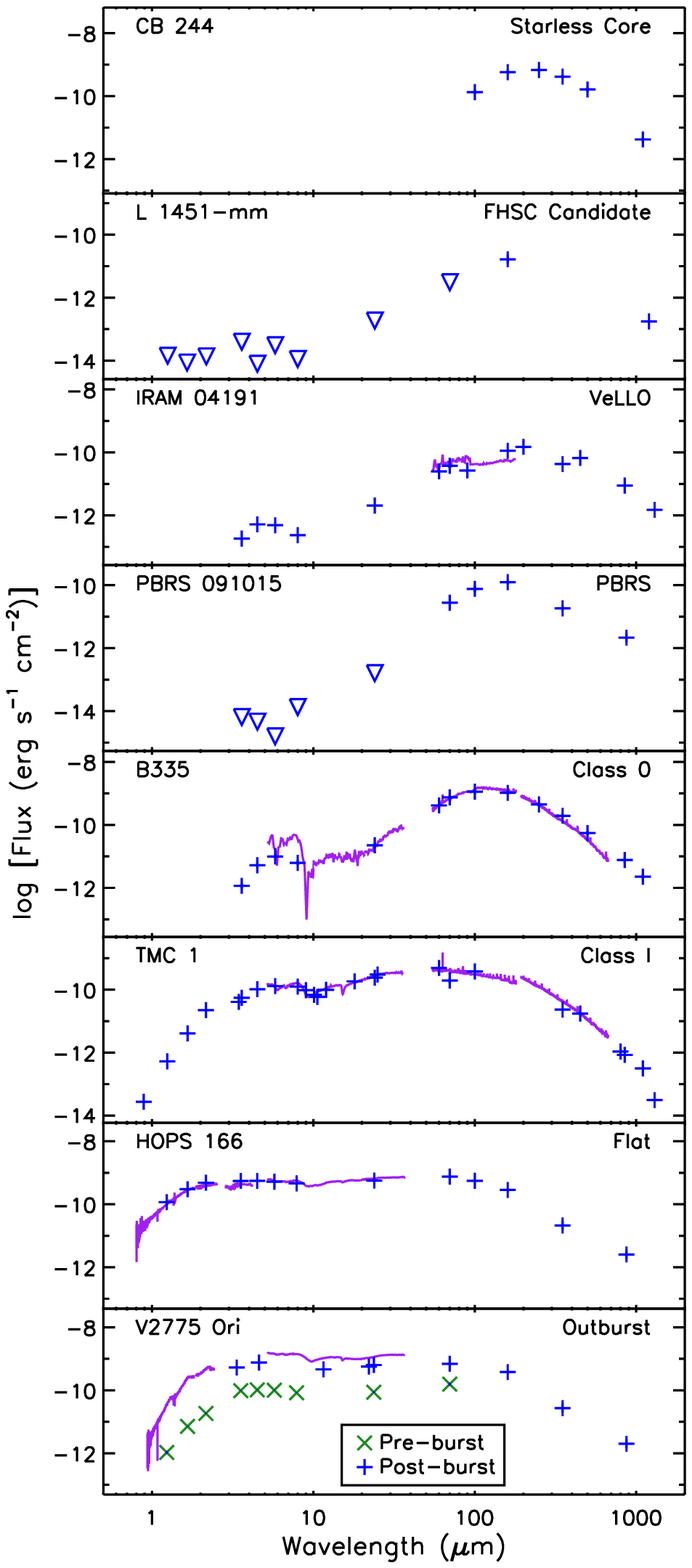}
\caption{\small SEDs for a starless core \citep{2010A&A...518L..87S,2013A&A...551A..98L}, a candidate first hydrostatic core \citep{2011ApJ...743..201P}, a very low-luminosity object \citep{2008ApJS..179..249D,2013ApJ...770..123G}, a PACS bright red source \citep{2013ApJ...767...36S}, a Class 0 protostar \citep{2008ApJ...687..389S,2013A&A...551A..98L,2013ApJ...770..123G}, a Class I protostar \citep{2013ApJ...770..123G}, a Flat-SED source \citep{2010A&A...518L.122F}, and an outbursting Class I protostar \citep{2012ApJ...756...99F}.  The $+$ and $\times$ symbols indicate photometry, triangles denote upper limits, and solid lines show spectra.
\label{f.sed}}
\end{figure}

Class 0 objects were later added as a fifth class for protostars too deeply 
embedded to detect in the near-infrared but inferred through other means 
\citep[i.e., outflow presence;][]{1993ApJ...406..122A}.  They were defined 
observationally as sources with a ratio of submillimeter to bolometric 
luminosity (\lsmmbol) greater than 0.5\%, where \lsmm\ is calculated 
for $\lambda \geq 350$ \um.  Fig.~\ref{f.sed} shows example SEDs for a starless core and for Class 0, Class I, and Flat-SED sources as well as several other types of objects that will be discussed in subsequent sections.  For the first time, {\it Spitzer} and {\it Herschel} surveys of star-forming regions have provided truly complete spectral coverage for most protostars.

Due to the effects of inclination, aspherical geometry, and foreground reddening, there is not 
a one-to-one correspondence between observational Class and physical Stage 
\citep[e.g.,][]{2003ApJ...591.1049W,2003ApJ...598.1079W,2006ApJS..167..256R,2008A&A...486..245C}.  To avoid confusion, \citet{2006ApJS..167..256R} proposed the use of Stages 0, I, II, and III when referring to the physical or evolutionary status of an object and the use of ``Class'' only when referring to observations.  We follow this recommendation throughout this review.  


While the original discriminant between Class 0 and I protostars is \lsmmbol, this quantity has historically been difficult to calculate because it 
requires accurate submillimeter photometry.  Another quantity often used in its 
place is the bolometric temperature (\tbol):  the effective temperature 
of a blackbody with the same flux-weighted mean frequency as the observed 
SED \citep{1993ApJ...413L..47M}.  \tbol\ begins near 20 K for deeply embedded protostars \citep{2013A&A...551A..98L} and eventually 
increases to the effective temperature of a low-mass star once all of the surrounding core and disk 
material has dissipated.  \citet{1995ApJ...445..377C} proposed the following 
Class boundaries in \tbol:  70 K (Class 0/I), 650 K (Class I/II), and 
2800 K (Class II/III).


With the sensitivity of {\it Spitzer}, Class 0 protostars are routinely detected in the infrared, and Class I sources by $\alpha$ are both Class 0 and I sources by \tbol\ \citep{2009ApJ...692..973E}.  Additionally, sources with flat $\alpha$ have $T_{\rm bol}$ consistent with Class I or Class II, extending roughly from 350 to 950 K, and sources with Class II and III $\alpha$ have $T_{\rm bol}$ consistent with Class II, implying that $T_{\rm bol}$ is a poor discriminator between $\alpha$-based Classes II and III \citep{2009ApJS..181..321E}.


\begin{figure}[t]
\epsscale{1.0}
\plotone{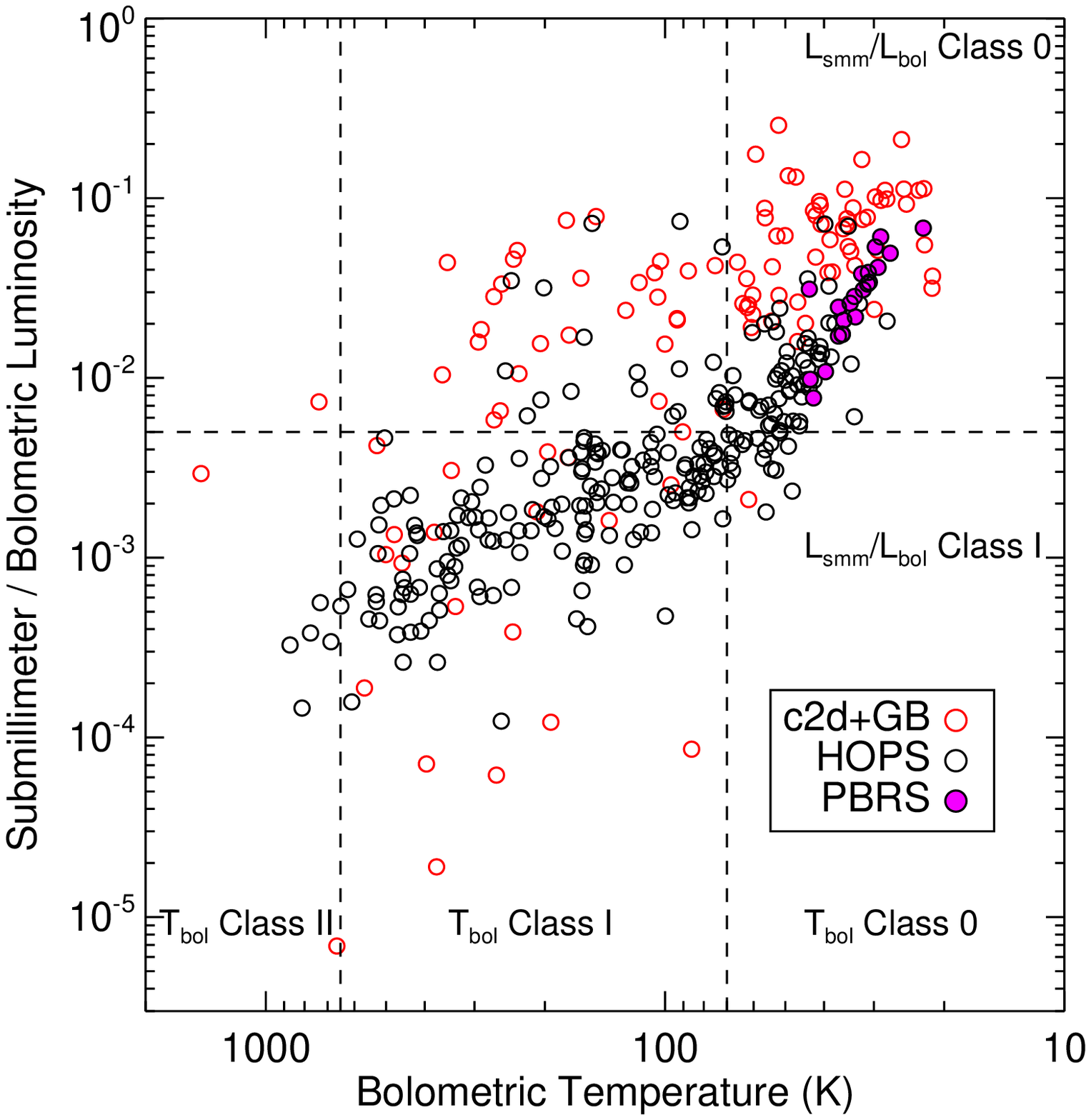}
\caption{\label{fig_classification}\small Comparison of \lsmmbol\ and \tbol\ for the protostars in the c2d, GB, and HOPS surveys.  The PBRS (\S \ref{sec_pbrs}) are the 18 Orion protostars that have the reddest 70 to 24 \micron\ colors,
11 of which were discovered with {\it Herschel}.  The dashed lines show the Class boundaries in \tbol\ from \citet{1995ApJ...445..377C} and in \lsmmbol\ from \citet{1993ApJ...406..122A}.  Protostars generally evolve from the upper right 
to the lower left, although the evolution may not be monotonic if accretion 
is episodic.}
\end{figure}

\tbol\ may increase by hundreds of K, crossing at least one Class boundary, as the inclination ranges from edge-on to pole-on \citep{2009A&A...507..861J,2013A&A...551A..98L,2013AN....334...53F}.  Thus, many Class 0 sources by \tbol\ may 
in fact be Stage I sources, and vice versa.  
Far-infrared and submillimeter diagnostics have a 
superior ability to reduce the influence of foreground reddening and 
inclination on the inferred protostellar properties.  At such wavelengths 
foreground extinction is sharply reduced and observations probe the colder, 
outer parts of the envelope that are less optically thick and thus where 
geometry is less important.  Flux ratios 
at $\lambda \geq 70$ \um\ respond 
primarily to envelope density, pointing to a means of disentangling these 
effects and developing more robust estimates of evolutionary stage 
\citep{2010A&A...518L.119A,2013ApJ...767...36S}.  
Along these lines, several authors have recently argued that \lsmmbol\ is 
a better tracer of underlying physical Stage than \tbol\ \citep{2005ApJ...627..293Y,2010ApJ...710..470D,2013A&A...551A..98L}.

Recent efforts have vastly expanded the available 
350 \um\ data for protostars via, e.g., the {\it Herschel} Gould Belt survey 
(see accompanying chapter by {\it Andr\'{e} et al.}), several {\it Herschel} 
key programs \citep[e.g.,][]{2013A&A...551A..98L,2013ApJ...770..123G}, and 
ground-based observations \citep[e.g.,][]{2007AJ....133.1560W}.  These efforts 
have enabled accurate calculation of \lsmmbol\ for large samples.  
Fig.~\ref{fig_classification} compares classification via \lsmmbol\ 
and \tbol\ for the c2d, GB, and HOPS protostars.  While there 
are methodology differences in the details of how \lsmm\ is calculated for 
the c2d+GB \citep{2013AJ....145...94D} and HOPS \citep{2013ApJ...767...36S} 
protostars that must be resolved in future studies, 
the two classification methods agree 
81\% of the time (counting \tbol\ Class II and \lsmmbol\ Class I as 
agreement).  In a similar analysis of nine isolated globules, 
\citet{2013A&A...551A..98L} did not find such a clear agreement between 
\tbol\ and \lsmmbol\ classification, although methodology and possibly 
environmental differences are substantial \citep{2013A&A...551A..98L}.  
Between the increased availability of submillimeter data and the evidence that 
\lsmmbol\ is a better tracer of underlying physical Stage, we suggest using 
\lsmmbol\ rather 
than \tbol\ as the primary tracer of the evolutionary status of protostars.  
However, the concept of monotonic evolution through the observational Classes 
breaks down if accretion is episodic (see \S \ref{sec_accretion}, 
\ref{sec_variability}).  Instead, protostars will move back and forth across 
class boundaries as their accretion rates and luminosities change \citep{2010ApJ...710..470D}.


Table \ref{t.nyso} lists the numbers of YSOs classified via $\alpha$ for the c2d+GB, Orion, and Taurus surveys.  Since portions of Orion suffer from incompleteness, we present data for subregions where the counts are most complete (L1630 and L1641).  Additionally, due to the high extinction toward Orion, we present for this region both the total number of Flat-SED sources as well as the number that are likely reddened Stage II objects based on analysis of longer-wavelength data.  Not included are new protostars in Orion discovered by {\it Herschel} (see \S \ref{sec_pbrs}); including these increases the total number by only 5\%--8\%, emphasizing that {\it Spitzer} surveys missed relatively few protostars.  

Table \ref{t.nyso} also gives lifetimes for the protostellar (Class 0+I) and Flat-SED objects.  These are calculated under the following set of assumptions: (1) time is the only variable, (2) star formation is continuous over at least the assumed Class II lifetime, and (3) the Class II lifetime is 2 Myr (see the accompanying chapter by {\it Soderblom et al.}).  This lifetime is best thought of as a ``half-life'' rather than an absolute lifetime.  Averaged over all surveys, we derive a protostellar lifetime of $\sim$0.5 Myr.  While the flat-SED lifetime by this analysis is $\sim$0.4 Myr, this class may be an inhomogeneous collection of objects that are not all YSOs at the end of envelope infall ({\S \ref{sec_timeline}).  In all cases these are lifetimes of {\it observed classes} rather than {\it physical stages}, since the latter are not easily observable quantities.  Indeed, some recent theoretical studies suggest the true lifetime of the protostellar stage may be shorter than the lifetime for the Class 0+I protostars derived here \citep{2011ApJ...736...53O,2012ApJ...747...52D}.

\begin{table}[t]
\caption{YSO Numbers and Lifetimes\label{t.nyso}}
\setlength{\tabcolsep}{4pt}
\begin{tabular}{lcccc}
\hline
\hline
                         & c2d+GB & L1630$^a$ & L1641$^b$ & Taurus \\
\hline
\multicolumn{5}{l}{Numbers} \\
Class 0+I$^c$            & 384      & 51     & 125    & 26       \\
Flat                     & 259      & 48     & 131    & 22       \\
Flat$^d$		 & $\cdots$ & 30     & 74     & $\cdots$ \\
Class II                 & 1413     & 243    & 559    & 125      \\
\hline
\multicolumn{5}{l}{Lifetimes$^e$} \\
Class 0+I (Myr)     & 0.54     & 0.42   & 0.45   & 0.42     \\
Flat (Myr)          & 0.37     & 0.40   & 0.47   & 0.35     \\
Class 0+I (Myr)$^f$ & $\cdots$ & 0.37   & 0.39   & $\cdots$ \\
Flat (Myr)$^f$      & $\cdots$ & 0.13   & 0.18   & $\cdots$ \\
\hline
\multicolumn{5}{l}{$^a$\small Omitting regions of high nebulosity}\\
\multicolumn{5}{l}{$^b$\small Omitting the Orion Nebula region}\\
\multicolumn{5}{l}{$^c$\small Not including new Herschel sources}\\
\multicolumn{5}{l}{$^d$\small Number of previous row that are likely reddened Class II}\\
\multicolumn{5}{l}{$^e$\small Assuming a Class II lifetime of 2 Myr (see text)}\\
\multicolumn{5}{l}{$^f$\small Counting sources in row marked $d$ as Class II}\\
\end{tabular}
\end{table}



\subsection{Protostellar Luminosities}

\begin{figure}[t]
\includegraphics[width=\hsize]{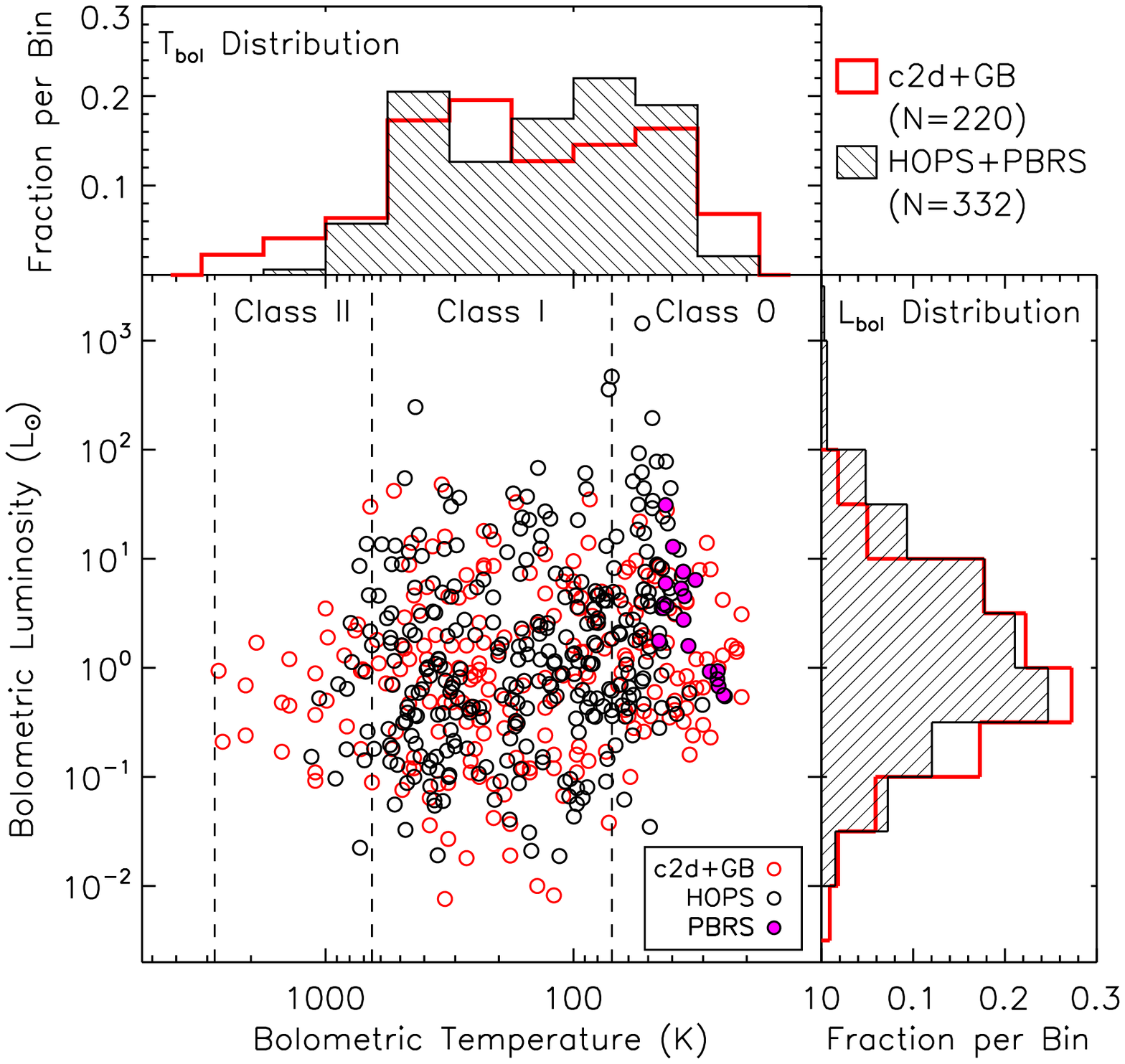}
\caption{\small The combined bolometric luminosity vs.\ temperature (BLT) diagram for the 552 protostars identified in the c2d, Gould Belt, and Spitzer Orion / HOPS surveys.  BLT properties are measured by integrating under the observed SEDs.  Marginal distributions of \tbol\ and \lbol\ are shown as histograms. The full population of protostars spans more than four orders of magnitude in luminosity.
\label{f.blt}}
\end{figure}

 Fig.~\ref{f.blt} plots \lbol\ vs.~\tbol\ \citep[a ``BLT'' diagram;][]{1993ApJ...413L..47M} for the protostars in the c2d+GB and Orion surveys, along with the separate \lbol\ and \tbol\ distributions.  The 220 c2d+GB protostars are the YSOs in those surveys with submillimeter detections.  
The 332 Orion protostars are the 317 identified by \citet{2012AJ....144..192M} that were detected in HOPS 70 \um\ observations, plus an additional 15 that were newly discovered by HOPS \citep[\S \ref{sec_pbrs}; ][]{2013ApJ...767...36S}.  Bolometric properties were calculated by trapezoidal integration under the available SEDs with no extrapolation to shorter or longer wavelengths and no corrections for foreground extinction.  As found in previous work \citep{1990AJ.....99..869K}, the luminosity distribution extends over several orders of magnitude, but this is now the case for hundreds of sources, and the distributions extend to even lower luminosities.  
Explanations for these broad distributions will be evaluated in \S \ref{sec_accretion}. 

Statistics for the \lbol\ distributions appear in Table~\ref{t.lum}.  Since the 
flux completeness limits of the c2d, GB, and HOPS surveys can be a function of 
position in regions with diffuse emission, and the 
relationship between source fluxes and \lbol\ depends on distance, source 
evolutionary status, local strength of the external radiation field, and total 
core mass available to be heated externally, it is difficult to derive an 
exact completeness limit in \lbol.  \citet{2013AJ....145...94D} derive an 
approximate completeness limit of 0.05 \lsun\ for all but the most distant 
cloud in the c2d and GB surveys (IC5146), which increases to 0.2 \lsun\ when 
IC5146 is included.  A full analysis of the HOPS completeness limit is still 
under investigation ({\it Stutz et al.}, in preparation).

The c2d+GB distribution extends over three orders of magnitude.   Except for a statistically significant  excess of low luminosity Class I sources, the distributions are generally similar for Classes 0 and I.  Looking at the same clouds, \citet{2012AJ....144...31K} find a distribution shifted to lower luminosities, with a mean (median) of 0.66 (0.63) \lsun.  The discrepancy between these two studies is partially resolved by updating the empirical relationship adopted by \citet{2012AJ....144...31K} to calculate \lbol, but some discrepancy remains \citep{2013AJ....145...94D}.  As discussed above, there are concerns with both the completeness of \citet{2013AJ....145...94D} and the reliability of \citet{2012AJ....144...31K}.  

Including new {\it Herschel} sources, the HOPS distribution extends over four orders of magnitude, with thirteen protostars that are more luminous than the most luminous c2d+GB source.  These sources increase the mean values of the HOPS luminosty distribution for both the Class 0 and I subsamples by factors of 6 and 2.4, respectively, and also increase the median Class 0 luminosity by a factor of 2.5 compared to the c2d+GB value.  The degree to which these changes can be attributed to environmental differences as opposed to incompleteness effects in the various samples 
is currently under investigation ({\it Stutz et al., in preparation}).


%
%
%
%
%

\begin{table}[t]
\caption{Protostellar \lbol\ Statistics (\lsun) \label{t.lum}}
\begin{tabular}{lcc}
\hline
\hline
 & c2d+GB & HOPS \\
\hline
Range & 0.01--69 & 0.02--1440 \\
Mean (Median) & 4.3 (1.3) & 14 (1.2) \\
Class 0 Mean (Median) & 4.5 (1.4) & 27 (3.5) \\
Class I Mean (Median) & 3.8 (1.0) & 9.3 (1.0) \\
\hline
\end{tabular}\\
{\small \sc note}: {\small Classes are by \tbol.  See text for details on which 
sources are counted as protostars and included here.}
\end{table}

\subsection{Refining the Timeline of Protostellar Evolution}\label{sec_timeline}

\begin{table}[t]
\caption{Protostellar Numbers by \tbol\ \label{t.c0}}
\begin{tabular}{lcc}
\hline
\hline
 & c2d+GB & HOPS \\
\hline
Class 0 & 63 & 93 \\
Class I & 132 & 222 \\
\hline
Fraction of Class 0$^a$ & 0.32 & 0.30 \\
Lifetime of Class 0 (Myr)$^b$ & 0.16 & 0.15 \\
\hline
\end{tabular}\\
$^{a}$ {\small Ratio of Class 0 to Class 0+I}\\
$^{b}$ {\small Ratio multiplied by the 0.5 Myr Class 0+I lifetime}\\
\end{table}

Determining when (and how quickly) envelopes dissipate is a major constraint 
for theories of how mass accretes onto 
protostars (see \S \ref{sec_accretion}).  One key metric is the relative 
numbers (and thus implied durations) of Class 0 and I protostars, which should 
be sensitive to the accretion history.  We list the numbers of each in Table 
\ref{t.c0}, using \tbol\ for classification since not all of the c2d and GB 
protostars have sufficient data for reliable \lsmmbol\ calculations.  
We find that 30\% of protostars are in Class 0 based on \tbol\ 
\citep{2009ApJ...692..973E,2013AJ....145...94D,2013AN....334...53F}, 
implying a Class 0 lifetime of 0.15 Myr, but 
interpreting these results is complicated by the lack of a one-to-one 
correspondence between Class and Stage.  
The combination of ubiquitous high-sensitivity 
far-infrared and submillimeter observations and improved modeling possible in 
the {\it Herschel} and ALMA eras will lead to a better understanding of 
the relative durations of these Stages.

In this review we have quoted a derived protostellar 
duration of $\sim$0.5 Myr.  However, the true duration remains uncertain.  
On one hand, 
some recent observations suggest that up to 50\% of Class I YSOs may be 
highly reddened Stage II objects 
\citep{2009A&A...498..167V,2010ApJ...723.1019H}, a possibility also discussed 
by \citet{2007prpl.conf..117W}.  However, the exact fraction is uncertain and 
may have been overestimated by these studies.  
On the other hand, the nature of the Flat-SED sources is 
unclear and has not yet been well-determined in the literature.  Some of them 
may represent objects in transition between Stages I and II 
\citep[e.g.,][]{1994ApJ...434..330C}, but determining 
their exact nature is critical for determining the duration of the 
protostellar stage.  Future work on this front is clearly needed.  
Finally, all the timescales discussed in this chapter scale directly with the 
timescale for the Class II phase, which we have assumed to be 2 Myr 
(see the accompanying chapters on the ages of young stars by 
{\it Soderblom et al.} and on transitional disks by {\it Espaillat et al.} for 
further discussion of this timescale).

\section{\textbf{PROTOSTELLAR ACCRETION}}\label{sec_accretion}


The fundamental problem of star formation is how stars accrete their mass.  In this section we give a general theoretical overview of how stars gain their masses and discuss the challenges of modeling protostellar luminosities. Understanding protostellar properties and evolution depends upon both the macrophysics of the core environment and the microphysics that drives gas behavior close to the protostar. After discussing protostellar luminosities (\S\ref{sec_lum}), we describe three categories of models: those that depend on core properties (\S\ref{core_reg}), those that are based on the core environment and feedback (\S\ref{feed_reg_acc}), and those focused on accretion disk evolution (\S\ref{disk_reg_acc}). This separation is mainly for clarity 
since actual protostellar accretion is determined by a variety of nonlinear and interconnected physical processes that span many orders of magnitude in density and scale.

\subsection{The Protostellar Luminosity Problem}\label{sec_lum}

The luminosity of a protostar provides an indirect measure of two quantities: the instantaneous accretion rate and the protostellar structure. Unfortunately, the luminosity contributions of each are difficult to disentangle.  
Over the past two decades various theories have attempted and failed to explain observed protostellar luminosites, which are found to be generally $\sim$10 times less luminous than expected \citep{1990AJ.....99..869K, 1995ApJS..101..117K,2005ApJ...627..293Y, 2009ApJS..181..321E}. 
This discrepancy became known as the ``protostellar luminosity problem'' and can be stated as follows.  The total protostellar luminosity is given by:
\beq\label{eq_luminosity}
L_p = L_{\rm phot} + f_{\rm acc} \frac{Gm \dot m}{r},
\eeq
where $L_{\rm phot}$ is the photospheric luminosity generated by deuterium burning and Kelvin-Helmholtz contraction, $m$ is the protostellar mass, $\dot m$ is the instantaneous accretion rate, $r$ is the protostellar radius, and $f_{\rm acc}$ is the fraction of energy radiated away in the accretion shock.
For accretion due to gravitational collapse (e.g., \citealt{1980ApJ...241..637S}), 
$\dot m \sim (c_s^2+v_A^2+v_t^2)^{3/2}/G \simeq 10^{-5}\, \msun$ yr$^{-1}$, where $c_s$, $v_A$, and $v_t$ are the sound, Alfv\'en, and turbulent speeds, respectively. Given a typical protostellar radius $r=3.0$ \rsun\ \citep{1988ApJ...332..804S, 1992ApJ...392..667P},
the accretion luminosity of a 0.25 \msun\ protostar is $\sim$ 25 \lsun. This is many times the observed median value and is only a lower limit to the true problem since it neglects the contributions from $L_{\rm phot}$ and from external heating by the interstellar radiation field.  

However, Eq.~\ref{eq_luminosity} contains several poorly constrained quantities. The accretion rate onto a protostar is due to a combination of infalling material driven by gravitational collapse on large scales and the transport of material through an accretion disk on small scales.  The disk properties are set by the core properties and the rate of infall \citep[e.g.,][see also the accompanying chapter by {\it Li et al.}]{2009ApJ...692.1609V,2010ApJ...708.1585K}, and significant theoretical debate continues on the relationship between infall through the core and accretion onto the protostar, both instantaneously and averaged over time.
Although it is difficult to directly measure $\dot m$, estimates for T Tauri stars based on infrared emission lines suggest $\dot m \lesssim 10^{-7}$ \msun\ yr$^{-1}$ \citep[e.g.,][]{1998AJ....116.2965M}.  Based on observational estimates for the embedded lifetime (see \S \ref{sec_surveys}), it is not possible for protostars to reach typical stellar masses without significantly higher average accretion during the embedded phase.  While similar measurements of infrared emission lines in the protostellar phase are especially difficult, some observations suggest similarly low rates for Class I objects \citep[e.g.,][]{1998AJ....116.2965M,2007prpl.conf..117W}.  Presently it is unclear if such observations are representative of all protostars or instead biased toward those objects already near the end of the protostellar stage and thus detectable in the near-infrared.  A higher accretion rate is inferred in at least one protostar from its current accretion luminosity and mass derived from the Keplerian velocity profile of its disk \citep{2012Natur.492...83T}.  Mass infall rates derived from millimeter molecular line observations are also typically higher \citep[e.g.,][]{2009A&A...502..199B,2013A&A...558A.126M}, although such rates are only available for some sources and are highly model-dependent. 

The intrinsic stellar parameters are also poorly constrained. While main-sequence stellar evolution is relatively well understood, pre-main sequence evolution, especially during the first few Myr, is much less well-constrained. Consequently, both $L_{\rm phot}$ and $r$ are uncertain and 
depend on the properties of the first core, past accretion history, accretion shock physics, and the physics of stellar interiors \citep{2009ApJ...702L..27B,2011ApJ...738..140H,2012ApJ...756..118B}.

The luminosity problem was first identified by \citet{1990AJ.....99..869K}, who also proposed several possible solutions, including slow and episodic accretion. In the slow accretion scenario, the main protostellar accretion phase lasts longer than a freefall time, which was then typically assumed to be $\sim$0.1 Myr. Current observations suggest a protostellar duration of $\sim0.5$ Myr (\S \ref{sec_surveys}), which helps to alleviate the problem. In the episodic accretion scenario, accretion is highly variable and much of the mass may be accreted in statistically rare bursts of high accretion, observational evidence for which is discussed in \S \ref{sec_variability} and in the accompanying chapter by {\it Audard et al.}  One additional solution concerns the radiative efficiency of the accretion shock. If some of the shock kinetic energy is absorbed by the star or harnessed to drive outflows such that $f_{\rm acc} < 1$ (e.g., \citealt{1995ApJ...447..813O}),
the radiated energy would be reduced. Recent models applying combinations of these solutions have made significant progress towards reconciling theory and observation, as discussed in the remainder of this Section.

\begin{figure}[t]
\plotone{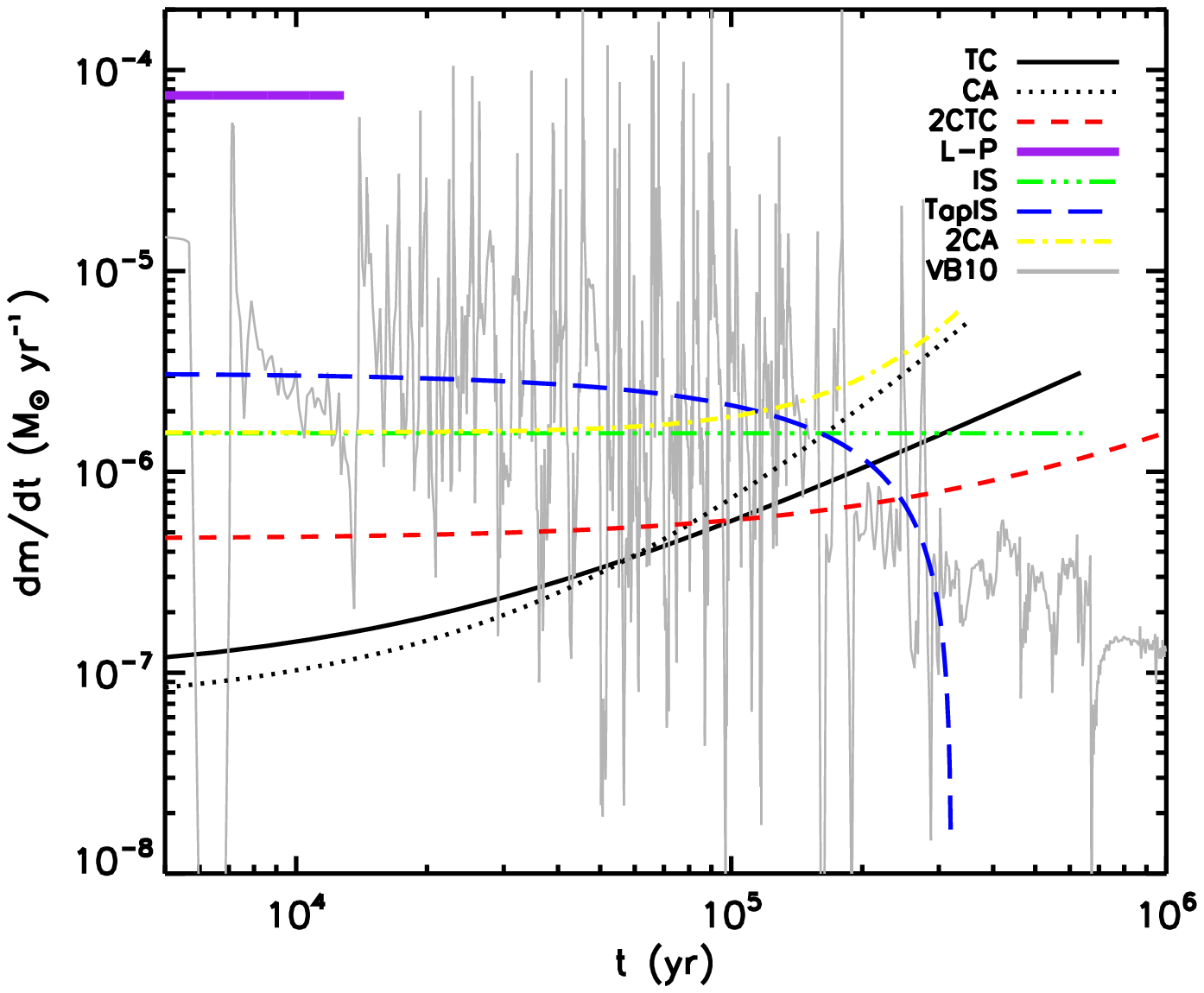}
\caption{\small Various model accretion rates as a function of time for a star of final mass 1 \msun. The models are the turbulent core (TC), competitive accretion (CA), two-component turbulent core (2CTC), Larson-Penston (L-P), isothermal sphere (IS), tapered isothermal sphere (TapIS), two-component accretion (2CA), and a simulated accretion history from \citet{2010ApJ...719.1896V}.  The TC, CA and 2CTC are computed assuming a mean formation time of $\avg{\tf}=0.44$ Myr.}\label{mdot}
\end{figure}

\subsection{Core-Regulated Accretion}\label{core_reg}

 A variety of theoretical models have been proposed for the gravitational collapse of dense cores. These models predict quantities such as the core density profile, gas velocities and, most crucially, the rate of mass infall to the core center. If the accretion rate of a forming protostar is identical to the gas infall rate, as expected in models where disks either efficiently transfer mass onto the protostar or do not form at all, then these models also predict protostellar accretion rates. Numerical simulations of forming clusters suggest that accretion rates predicted by theoretical models are on average comparable to the stellar accretion rate \citep[e.g.,][]{2012ApJ...754...71K}.  In this section and in \S \ref{feed_reg_acc} we discuss models that assume that the accreting gas is efficiently channeled from 0.1 pc to 0.1 AU scales, equivalent to focusing on the time-averaged accretion rate.  Models where the instantaneous and time-averaged accretion rates may diverge significantly are discussed in \S \ref{disk_reg_acc}.


Fig.~\ref{mdot} illustrates the accretion rate versus time for a 1 \msun\ star for a variety of theoretical models.  Most core-regulated accretion models are based on the assumption that gas collapses from a local dense reservoir of order $\sim$0.1 pc, i.e., a ``core''. The collapse of an isothermal, constant density sphere including thermal pressure is the simplest model with a self-similar solution. 
\citet{1969MNRAS.145..271L} and \citet{1969MNRAS.144..425P} separately calculated the resulting accretion rate to be $\dot m = 46.9 c_{\rm s}^3/G = 7.4 \times 10^{-5} (T/{\rm 10 K})^{3/2}\msun$ yr$^{-1}$, where $c_s$ and $T$ are the thermal sound speed and temperature, respectively. The self-similar solution for the collapse of a centrally condensed, isothermal sphere was computed by \citet{1977ApJ...214..488S}, 
who found $\dot m = 0.975 c_{\rm s}^3/G$, a factor of $\sim$50 less than the Larson-Penston solution. 
In both cases, accretion does not depend on the initial mass, which leads to the prediction that accretion rate is independent of the instantaneous protostellar and final stellar mass and implies that the only environmental variable that affects accretion is the local gas temperature.

The above models consider only thermal pressure and gravity.  However, cores are observed to be both magnetized and somewhat turbulent. The most massive clumps (e.g., \citealt{2011ApJS..196...12B}),
which are possible progenitors of massive protostars, have turbulent linewidths several times the thermal linewidth. 
\citet{2003ApJ...585..850M} proposed a ``turbulent core'' model to account for the higher column densities and turbulent linewidths of high-mass cores. 
In this model, $\dot m \propto m^{1/2} \mf^{3/4}$, where $m$ and $\mf$ are the instantaneous and final protostellar masses, respectively.

Significant numerical work has been devoted to modeling the formation of star clusters (see the accompanying chapter by {\it Offner et al}).  Many of these simulations present a very dynamical picture in which protostellar accretion rates vary as a function of location within the global gravitational potential and protostars ``compete'' with one another for gas \citep{2001MNRAS.323..785B}. 
The most massive stars form in the center of the gravitational potential where they can accrete at high rates.  
Protostars accrete until the gas is either completely accreted or dispersed, leading to a constant accretion time for all stars that is proportional to the global freefall time. Analytically, this suggests $\dot m \propto m^{2/3}\mf$ \citep{2001MNRAS.323..785B,2010ApJ...716..167M}.

All core-regulated accretion models then fall somewhere between the limits of constant accretion rate and constant star formation time.  
\citet{2010ApJ...716..167M} propose hybrid models that include both a turbulent and a thermal component (``two-component turbulent core'') or a competitive and a thermal component (``two-component competitive accretion''). Additional models have sought to analytically include rotation \citep{1984ApJ...286..529T}, 
nonzero initial velocities \citep{2004ApJ...615..813F}, 
and magnetic fields \citep{2007ApJ...671..497A}. 
For example, \citet{2007ApJ...671..497A} 
modify the isothermal sphere collapse problem to consider ambipolar diffusion. They derive accretion rates that are enhanced by a factor of 2-3 relative to the non-magnetized case. 

Declining accretion rates that fall by an order of magnitude or more from their peak are produced in some numerical simulations \citep{2008ApJ...676L.139V,2009ApJ...703..131O}. 
\citet{2010ApJ...716..167M} model this decline by imposing a ``tapering'' factor, $(1-t/\tf)$, such that accretion declines and terminates at some specified formation time, $\tf$.  
%
%
Infall could also be variable due to 
turbulence or magnetic effects \citep[e.g.,][]{2005ApJ...618..783T}, but 
such scenarios have not been well studied theoretically and are 
observationally difficult to constrain.

\subsection{Feedback-Regulated Accretion}\label{feed_reg_acc}

Most stars are born in cores in extended molecular clouds that are surrounded by lower-density filamentary gas and other recently formed protostars \citep[e.g.,][]{2007ARA&A..45..339B,2009ApJS..181..321E}.  These core environments provide mass for accretion from beyond the core, and  ``stellar feedback'' in the form of ionizing radiation, winds and outflows may also disperse star-forming gas. In this case the core environment is important, and the initial core mass is not sufficient to predict the final protostar mass.  Instead the protostar mass depends on its accretion history and the competition between infall and dispersal. 

Early models of such feedback-regulated accretion 
concentrated on the competition between accretion and outflows \citep{1980ApJ...238..158N}. 
\citet{2004MNRAS.347L..47B} proposed a lognormal distribution of initial core masses, whose accretion rates are proportional to their mass, and accretion durations follow a waiting-time distribution.  For this distribution, the probability density that accretion endures for $t$ and then stops between $t$ and $t+ dt$  is  $(1/\tau)e^{-t/\tau}$, where $\tau$  is the mean accretion duration. A similar distribution of accretion durations was proposed to describe ejections by small multiple systems \citep{2005MNRAS.356.1201B}.

Recently, \citet{2010ApJ...714.1280M} proposed a feedback-regulated model that accounts for protostellar masses that follow the IMF.
The basic ideas of the model are (1) protostars accrete from core-clump condensations, (2) the duration of accretion is the most important factor in setting the final mass of a protostar, and (3) accretion durations vary due to a combination of ejections, dispersal by stellar feedback, accretion competition, and exhaustion of initial gas.  The waiting-time distribution describes this combination.

    In dense clusters, a promising direction for protostar mass models is to treat the mass accretion rate as a function of mass or time, rather than as due to the collapse of a particular initial configuration.  This approach allows discrimination between various IMF-forming accretion models 
(see also \S \ref{core_reg}). 
\citet{2009ApJ...706.1341M,2010ApJ...714.1280M} formulate a two-component accretion rate denoted ``2CA'' having a constant, thermal component and a mass-dependent component (see Fig.~\ref{mdot}). 
The 2CA accretion rate is similar to that of the two-component turbulent core model. \citet{2010ApJ...714.1280M} combined this 2CA model with an explicit distribution of accretion durations to derive the protostellar mass distribution and showed that it closely resembles the stellar IMF.


\subsection{Disk-Regulated Accretion}\label{disk_reg_acc}


Since most core mass likely accretes through a disk 
(\S \ref{sec_disks}), it is important to determine how
this mass is redistributed within the disk and transported 
onto the star. Two main processes of mass and angular momentum transport in 
disks have been proposed: viscous torques due to turbulence 
triggered by the magneto-rotational instability (MRI) \citep{1991ApJ...376..214B} 
and gravitational torques induced by gravitational instability (GI) 
\citep{1987MNRAS.225..607L,1994ApJ...436..335L}. 
Which of these mechanisms dominates and the efficiency with which the disk 
transports mass onto the protostar (and thus the degree to which the 
instantaneous infall and accretion rates are coupled) are currently debated 
and may depend on the details of local physical conditions, including the core 
properties and the infall rate.  
Numerical models that circumvent the complicated physics of the MRI
and GI and treat both as a local viscous transport mechanism have been
developed based on the \citet{1973A&A....24..337S} $\alpha$-parameterization.
These $\alpha$-disk models have been successful in describing many aspects
of disk physics and accretion 
\citep[e.g.,][]{2008ApJ...681..375K,2009ApJ...694.1045Z}, 
though their applicability may be limited for fairly massive disks
\citep{2010NewA...15...24V}.

The MRI is known to operate only if the ionization fraction is sufficiently 
high to couple the magnetic field with the gas \citep{1994ApJ...421..163B}. 
Protostellar disks are sufficiently cold and dense that known ionization sources (thermal, X-rays, 
cosmic rays) may fail to provide the ionization needed to sustain the 
MRI, particularly in the disk mid-plane. If this is the case, \citet{1996ApJ...457..355G} proposed that the MRI may be active only in a layer near the surface, an idea known as ``layered accretion''.
Various disk models 
suggest that the so-called dead zone, wherein the MRI is largely suppressed, may 
occupy a significant fraction of the disk volume at AU scales 
\citep[see review by][]{2011ARA&A..49..195A}.

On the other hand, analytic models and numerical hydrodynamics simulations 
indicate that the physical conditions in protostellar disks may be 
favorable for the development of GI \citep{1964ApJ...139.1217T, 1987MNRAS.225..607L, 
1994ApJ...436..335L}. Whenever the destabilizing effect of self-gravity 
becomes comparable to 
the stabilizing effects of pressure and 
shear, spiral density waves develop. 
If heating due to GI is balanced 
by disk cooling, the disk settles into a quasi-steady state in which GI transports 
angular momentum outwards allowing mass to accrete onto the central star 
\citep{2004MNRAS.351..630L}. In this picture, gravitational torques alone
are sufficient to drive accretion rates consistent 
with observations of intermediate- and upper-mass T Tauri stars 
\citep{2007MNRAS.381.1009V,2008ApJ...676L.139V}.
However, for the very low-mass disks around low-mass stars and brown dwarfs, 
GI is likely to be suppressed and, consequently, viscous mass transport
due to MRI alone may explain the accretion rates of these objects \citep{2009ApJ...703..922V}.

The degree of GI depends on a number of factors including the angular momentum of the parent core, infall rate, disk mass, and amount of radiative heating from the central protostar \citep{2010ApJ...708.1585K,2010ApJ...725.1485O,2010ApJ...719.1896V,2011ApJ...730...32S}.
If the disk is relatively massive and the local disk cooling time is faster than the dynamical time, sections of 
spiral arms can collapse into bound fragments 
and lead to a qualitatively different mode of disk evolution. In this mode, disk
accretion is an intrinsically variable process due to disk fragmentation, 
nonaxisymmetric structure, and 
gravitational torques \citep{2009ApJ...704..715V}.
Fragments that form in the disk outer regions are quickly 
driven inwards due to the loss of angular momentum via 
gravitational interaction with the spiral arms 
\citep{2005ApJ...633L.137V,2006ApJ...650..956V,2011MNRAS.416.1971B,2011MNRAS.415.3319C}. 
As they accrete onto the protostar, clumps trigger luminosity bursts similar in magnitude to FU-Orionis-type or EX-Lupi-type 
events \citep{2005ApJ...633L.137V,2006ApJ...650..956V,2010ApJ...719.1896V,2011ApJ...729...42M}.

This burst mode of accretion mostly operates in the embedded phase of 
protostellar evolution when the continuing infall of gas from the parent core 
triggers repetitive episodes of disk fragmentation. The mass accretion onto the 
burgeoning protostar is characterized by short ($\la 100-200$~yr) bursts 
with accretion rate $\dot{m} \ga \mathrm{a~few} \times 10^{-5} M_\odot$~yr$^{-1}$ 
alternated with longer ($10^3-10^4$~yr) quiescent periods with $\dot{m}\la 10^{-6} 
M_\odot$~yr$^{-1}$ \citep{2010ApJ...719.1896V}. After the parent core is accreted or dispersed, any remaining 
fragments may trigger final accretion bursts 
or survive to form wide-separation planets or brown dwarfs 
\citep{2010ApJ...714L.133V,2010ApJ...710.1375K,2013A&A...552A.129V}.

Gravitational fragmentation is one of many possible mechanisms that can generate accretion and luminosity bursts (see accompanying chapter by {\it Audard et al.}).  
Among the other mechanisms, a combination of MRI and GI  
has been most well-studied 
\citep{2009ApJ...694.1045Z,2010ApJ...713.1134Z,2012MNRAS.423.2718M}. 
In this scenario, GI in the outer disk 
transfers gas to the inner sub-AU region where it accumulates until 
the gas density and temperature reach values sufficient for thermal ionization to activate the MRI.
The subsequent enhanced angular momentum transport 
triggers a burst of accretion.
The GI+MRI burst mechanism may act in disks that are not sufficiently massive
to trigger disk fragmentation, but the details of the MRI are still poorly understood.
Independent of their physical origin, luminosity bursts have important implications for disk fragmentation, accretion, and theoretical models of star formation.

\subsection{Comparison between Models and Observations}\label{theor_comp}

Here, we consider direct comparisons between observations and several theoretical models.

\subsubsection{Protostellar Luminosities}

In order to consider the luminosity problem, \citet{2010ApJ...716..167M} 
developed an analytic formalism for the present-day protostellar mass function (PMF).  
The PMF depends on the instantaneous protostellar mass, final mass, accretion rate, and average protostellar lifetime. Given some model for protostellar luminosity as a function of protostellar properties, \citet{2011ApJ...736...53O} then derived the present-day protostellar luminosity function (PLF). Since the PLF depends only on observable quantities such as the protostellar lifetime and on a given theoretical model for accretion, the PLF can be used to directly compare star formation theories with observations.


\citet{2011ApJ...736...53O} computed the predicted PLFs for a variety of models and parameters, including the isothermal sphere, turbulent core, competitive accretion, and two-component turbulent core models. 
Fig.~\ref{lum} compares the observed protostellar luminosity distribution with some of these predicted PLFs where the accretion rate is allowed to taper off as the protostar approaches its final mass.  Models in which the accretion rate depends on the final mass (such as the turbulent core or competitive accretion models) naturally produce a broad distribution of luminosities. \citet{2011ApJ...736...53O} also found that the theoretical models actually produce luminosities that are {\it too dim} compared to observations, given an average formation time of $\avg{\tf}\sim0.5$ Myr and allowing for episodic accretion. They concluded that a star formation time of $\avg{\tf} \simeq 0.3$ Myr provides a better match to the mean and median observed luminosities.


On the other hand, given that numerical simulations of disk evolution indicate that protostellar accretion may be an intrinsically variable process, the wide spread in the observed protostellar luminosity distribution may result from large-scale variations in the protostellar accretion rate \citep{2010ApJ...710..470D}. This idea was further developed by 
\citet{2012ApJ...747...52D}, who used numerical hydrodynamics simulations of 
collapsing cores coupled with radiative transfer calculations to compare the model and observed 
properties of young embedded sources in the c2d clouds. 
They showed that gravitationally unstable disks with accretion rates that both decline with time
and feature short-term variability and episodic bursts can 
reproduce the full spread of observations, 
including very low luminosity objects. As shown in Fig. \ref{lum}, accretion variability induced by GI and disk 
fragmentation can thus
provide a reasonable match to the observed protostellar luminosity distribution and resolve the
long-standing luminosity problem.

Finally, the distribution of protostellar masses can be obtained for the case of feedback-regulated accretion by combining the two-component accretion (2CA) model with an explicit distribution of accretion times.  \citet{2011ApJ...735...82M,2012ApJ...752....9M} used the corresponding distribution of masses and accretion rates to compute the predicted PLF and found that it is in reasonable agreement with the observed protostellar luminosity distribution in nearby clouds.

\begin{figure}[t]
\epsscale{1.05}
\plotone{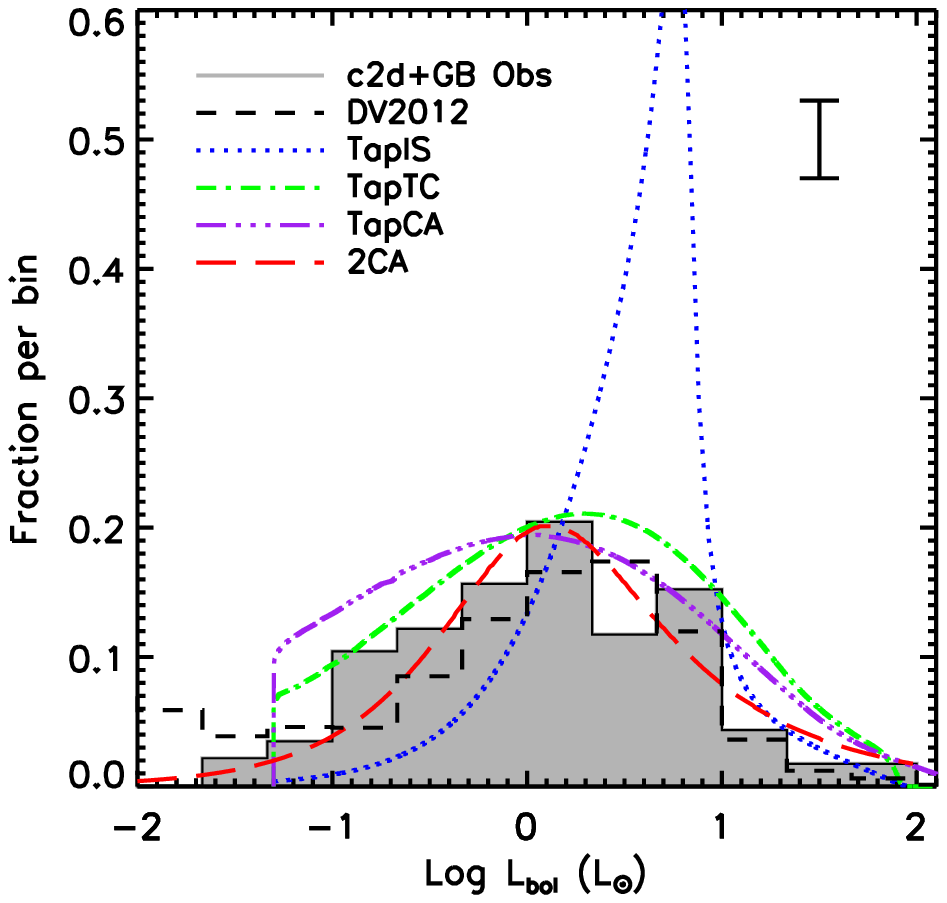}
\vspace{-0.2in}
\caption{\small Distribution of extinction-corrected protostellar luminosities from the c2d+GB surveys (shaded), predicted from disk simulations \citep[dashed,][]{2012ApJ...747...52D}, tapered Isothermal Sphere, tapered Turbulent Core, tapered Competitive Accretion \citep{2011ApJ...736...53O}, and two-component accretion \citep{2012ApJ...752....9M}. The tapered models adopt a completeness limit of 0.05 $\lsun$. Typical observational uncertainties are shown in the upper right.}\label{lum}
\end{figure}

\subsubsection{Ages of Young Clusters}



Models which specify accretion durations can also be tested against age estimates of young clusters. Such models can predict as a function of time the number of cluster members which are protostars, since they are still accreting, and the number which are pre-main sequence stars (PMS), since they have stopped accreting.  Application of the 2CA model to embedded cluster members identified as protostars or PMS indicate typical cluster ages of 1-3 Myr, in good agreement with estimates from optical and infrared spectroscopy and pre-main sequence evolutionary tracks.  This method can be used to date obscured young subclusters that are inaccessible to optical spectroscopy \citep{2012ApJ...752....9M}.

\subsubsection{Future Work}

The models described above provide tangible, diverse resolutions to the luminosity problem, including mass-dependent and highly-variable accretion histories.  Future observational work should concentrate on constraining the magnitude and timescales of protostellar variability to assess its effects on protostellar luminosities and, ultimately, its importance in the mass accretion process.  On the theoretical front, high resolution global-disk simulations, which include protostellar heating, magnetization, and ionization are needed to improve our understanding of disk-regulated accretion variability. Improved theoretical understanding of outflow launching and evolution is needed to interpret observed outflow variability and how it correlates with accretion. 
Additional synthetic observations of models that take into account the complex, asymmetric morphologies of accreting protostars and the effects of external heating are required to properly compare to observed luminosities.  Finally, dust evolution and chemical reaction networks determine the distribution of species, which is the lens through which we perceive all observational results. Many theoretical studies gloss over or adopt simplistic chemical assumptions, which undermines direct comparison between theory and observation.

\section{\textbf{THE EARLIEST OBSERVABLE STAGES OF PROTOSTELLAR EVOLUTION}}\label{sec_earliest}

The identification of young stars in the very earliest stages of formation 
is motivated by the goals of studying the initial conditions of the dense 
gas associated with star formation before modification by feedback 
and of determining the properties of cores when collapse begins.  The earliest
theoretically predicted phases are the first hydrostatic core and Stage 0 
protostars.  Observational classes of young objects include
Class~0 sources, very low luminosity objects (VeLLOs; \S \ref{sec_vellos}), 
candidate first hydrostatic cores (FHSC; \S \ref{sec_fhsc}), and PACS bright 
red sources (PBRS; \S \ref{sec_pbrs}).  The challenge is to 
unambiguously tie observations to theory, a process complicated by optical 
depth effects, geometric (inclination) degeneracies, and
intrinsically faint and/or very red observed SEDs.
We begin with a theoretical overview of the earliest stages 
of evolution and then follow with observational anchors provided by 
\spitzer, \herschel, and other facilities.

\subsection{\textbf{Theoretical framework}}

Once a dense molecular cloud core begins to collapse, the earliest 
object that forms is the first hydrostatic core (first core, FHSC).
First predicted by
\citet{1969MNRAS.145..271L}, the FHSC exists between the starless and
protostellar stages of star formation and has not yet been
unambiguously identified by observations.  The FHSC forms once the 
central density increases to the point where the
inner region becomes opaque to radiation
\citep[$\rho_c \ga 10^{-13}$~g~cm$^{-3}$;][]{1969MNRAS.145..271L},
rendering the collapse adiabatic rather than isothermal.  This object
continues to accrete from the surrounding core and both its mass and
central temperature increase with time.  Once the temperature reaches
$\sim$~2000~K, the gravitational energy liberated by accreting material 
dissociates H$_2$, preventing the temperature from continuing to rise to
sufficiently balance gravity.  At this point the second collapse is
initiated, leading to the formation of the second hydrostatic core,
more commonly referred to as the protostar.

\begin{table}[t]
\caption{Predicted properties of FHSCs \label{tab:fhsc}}
\begin{tabular}{lcl}
\hline
\hline
Property & Range & References\\
\hline
Maximum Mass [\msun]           & $0.04-0.05$                     &
1, 2, 3, 4 \\
                            & $0.01-0.1^{a}$                      & 5, 6 \\
Lifetime [kyr]               & $0.5 - 50$  &
1, 3, 4, 6, 7 \\
Internal Luminosity [\lsun]    & $10^{-4} - 10^{-1}$                    & 2, 3 \\
Radius [AU]                 & $\sim 5$                              & 2 \\
                            & $\ga 10-20^{a}$                      & 5, 6 \\
\hline
\end{tabular}
$^{a}$ {\small With the effects of rotation included.}\\
{\small References: (1) \citet{1995ApJ...439L..55B}; 
\,(2) \citet{1998ApJ...495..346M}; 
\,(3) \citet{2007PASJ...59..589O}; 
\,(4) \citet{2010ApJ...725L.239T}; 
\,(5) \citet{2006ApJ...645..381S}; 
\,(6) \citet{2008ApJ...674..997S}; 
\,(7) \citet{2012A&A...545A..98C}  
}
\end{table}

The current range in predicted FHSC properties are
summarized in Table~\ref{tab:fhsc}.  The significant variation in
predicted properties is largely driven by different prescriptions and
assumptions about magnetic fields \citep[e.g.,][]{2012A&A...545A..98C},
rotation \citep[e.g.,][]{2006ApJ...645..381S,2008ApJ...674..997S}, and
accretion rates.  
Rotation may produce a flattened, disk--like
morphology for the entire FHSC 
\citep{2006ApJ...645..381S,2008ApJ...674..997S}, implying that disks may 
actually form before protostars with masses larger than the protostars 
themselves
\citep{2011MNRAS.417.2036B,2011MNRAS.413.2767M}.  

Studies of FHSC SEDs have shown that the emitted radiation is
completely reprocessed by the surrounding envelope, with SEDs
characterized by emission from $10-30$~K dust and no observable
emission below $\sim$$20-50$~\um\
\citep{1995ApJ...439L..55B,1998ApJ...495..346M,2007PASJ...59..589O,2011ApJ...728...78S}.
Emission profiles in various molecular species and simulated ALMA continuum images 
\citep{2011ApJ...728...78S,2011PASJ...63.1151T,2012A&A...548A..39C,2012ApJ...760...40A}
demonstrate the critical value of additional observational constraints beyond 
continuum SEDS.  
\citet{2008ApJ...676.1088M} showed that first cores drive slow
($\sim$\,5~\kms) outflows with wide opening--angles while 
\citet{2012MNRAS.423L..45P} showed that, under certain 
assumptions, first core outflows may show strong
collimation. Future theoretical attention is needed.  Indeed, despite
the dire observational need, no clear and unambiguous predictions for
how a first core can be differentiated from a very young protostar
have yet been presented.

\subsection{\textbf{Observations}}

Here we outline recent observational developments in the
identification of the youngest sources, focusing on discoveries of
three classes of observationally defined objects mentioned above:
Very low luminosity objects (VeLLOs), candidate FHSCs, and PACS
bright red sources (PBRS).

\subsubsection{\textbf{Very Low Luminosity Objects (VeLLOs)}}\label{sec_vellos}

Prior to the launch of {\it Spitzer}, dense cores were identified 
as protostellar or starless based on the presence or absence of an associated 
{\it IRAS} detection.  Beginning with 
\citet{2004ApJS..154..396Y}, {\it Spitzer} c2d observations revealed a number 
of faint protostars in cores originally classified as starless.  This led to 
the definition of a new class of objects called very low luminosity objects 
(VeLLOs): protostars embedded in dense cores with internal luminosities 
\lint\ $\leq$ 0.1 \lsun\ 
\citep{2007prpl.conf...17D}, where \lint\ excludes the luminosity
arising from external heating by the interstellar radiation field. 
A total of 15 VeLLOs have been identified in the c2d regions 
\citep{2008ApJS..179..249D}, 
with six the subject of detailed observational and modeling studies 
\citep{2004ApJS..154..396Y,2006ApJ...651..945D,2006ApJ...649L..37B,2009ApJ...693.1290L,2010ApJ...721..995D,2011MNRAS.416.2341K}. 

The above studies have postulated three explanations for VeLLOs, whose
very low luminosities require very low protostellar masses and/or
accretion rates: (1) Extremely young protostars with very little mass
yet accreted, (2) Older protostars observed in quiescent periods of a
cycle of episodic accretion, and (3) Proto-brown dwarfs.  The
properties of the host cores and outflows driven by VeLLOs vary
greatly from source to source, suggesting that VeLLOs (which are
defined observationally) do not correspond to a single evolutionary
stage and are instead a heterogeneous mixture of all three
possibilities listed above.  While the relatively strong mid-infrared detections
guarantee that none are first hydrostatic cores, at least some VeLLOs
are consistent with being extremely young Class 0 protostars just beyond the 
end of the first core stage (see \S \ref{sec_fhsc}).

\subsubsection{\textbf{Candidate First Hydrostatic Cores}}\label{sec_fhsc}

Nine other objects embedded within cores originally classified as starless 
have been detected and identified as candidate first cores.  These objects have 
been revealed through faint mid--infrared detections of compact 
sources below the sensitivities of the large {\it Spitzer} surveys 
\citep{2006A&A...454L..51B,2010ApJ...722L..33E}, 
(sub)millimeter detections of molecular outflows driven by ``starless'' cores 
\citep{2010ApJ...715.1344C,2011ApJ...743..201P,2012ApJ...745...18S,2012ApJ...751...89C,2013ApJ...764L..15M},
and far--infrared detections indicating the presence of warm dust
heated by an internal source
\citep{2012A&A...547A..54P}.  However, we caution that even the very 
existence of some of these objects remains under debate 
(e.g., {\it Schmalzl et al.} 2013, in prep.).

\begin{figure*}
 \begin{center} \scalebox{0.43}{\includegraphics{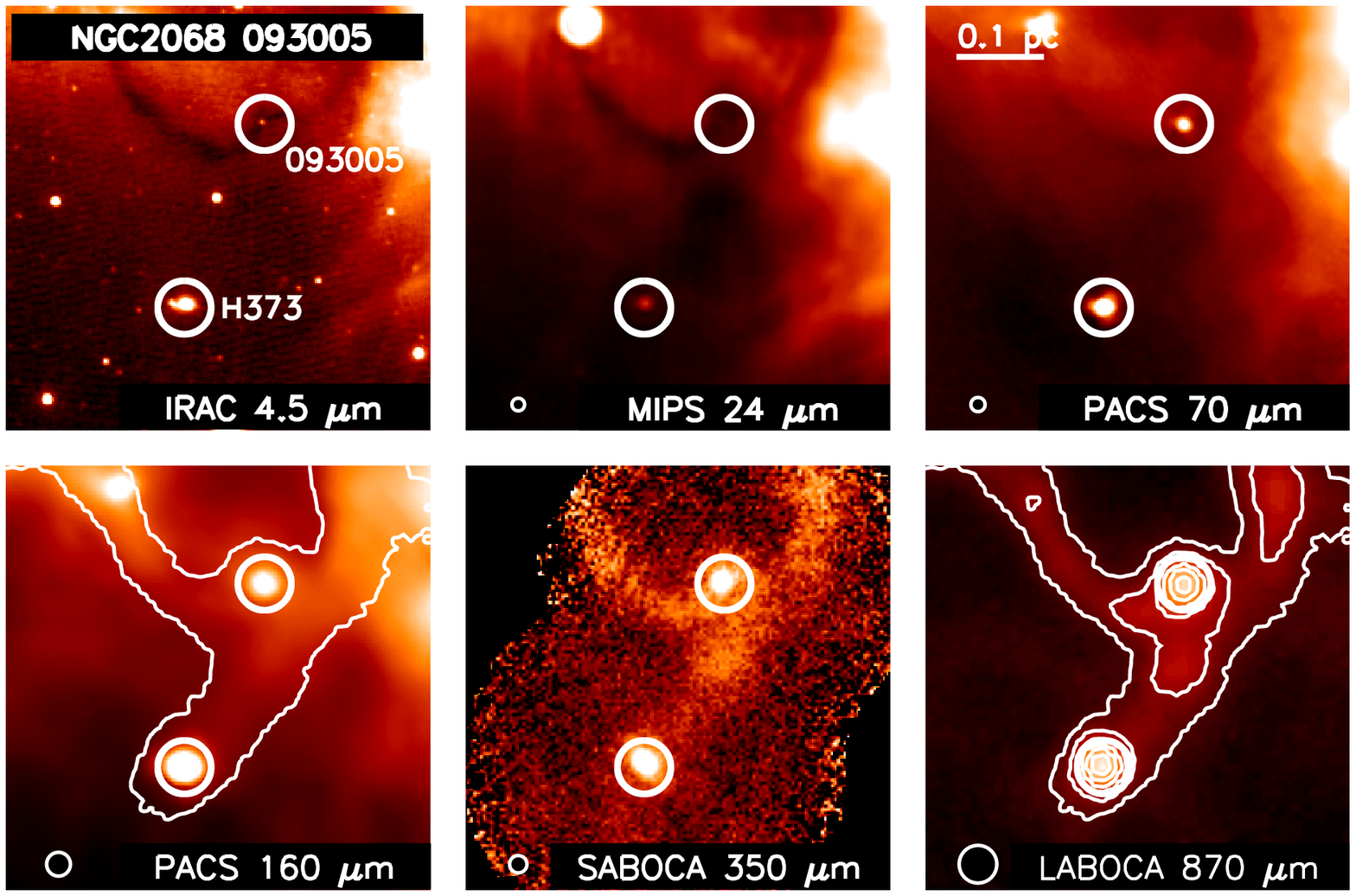}\includegraphics{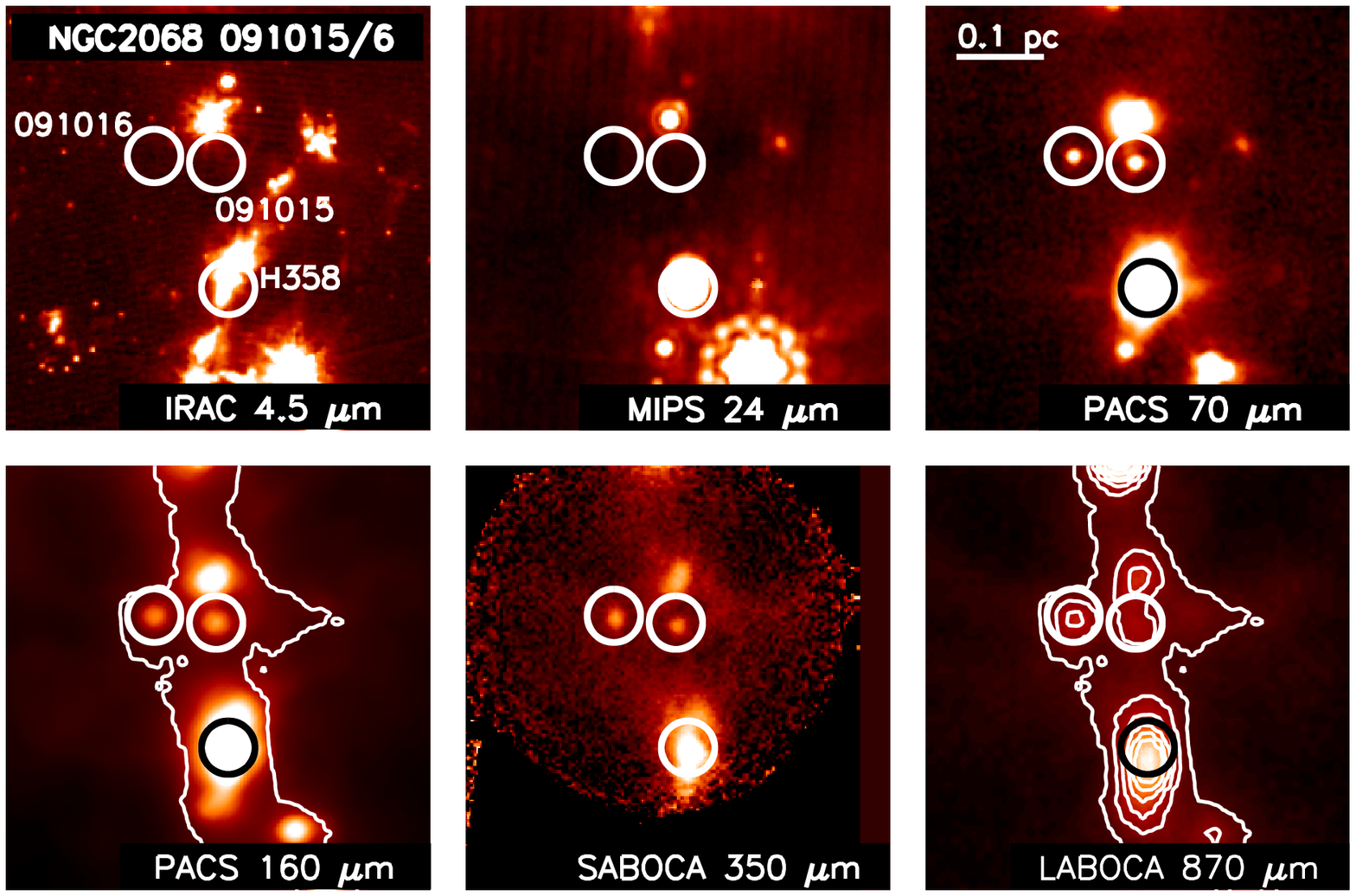}} \caption{\small
 $4\arcmin \times 4\arcmin$ images of 5 PBRS at the indicated
 wavelengths.  {\it Left:} PBRS 093005 and H373. Source 093005 is the
 reddest PBRS and therefore the reddest embedded object known in
 Orion, and lies at the intersection of three filaments seen in both
 absorption (8, 24, and 70~\um) and in emission (350 and 870~\um).
 {\it Right:} PBRS 091015, 019016, and H358.  Figure adapted
 from \citet{2013ApJ...767...36S}.}  \label{fig:stutz_01} \end{center}
\end{figure*} 

Two significant questions have emerged: 1) Are any candidates
true first cores?  2) How many ``starless'' cores are truly starless?
The answer to the first question is not yet known; none clearly stand
out as the best candidate(s) for a bona--fide FHSC, with arguments for
and against each.  Nonetheless, it is extremely unlikely that all are FHSC.  
Six of the nine candidate FHSCs are located in the Perseus molecular cloud.  
\citet{2009ApJ...692..973E} identified 66 protostars in Perseus.
Assuming that the duration of the protostellar stage is $\sim$0.5 Myr
(\S \ref{sec_surveys}) and a first core lifetime of $0.5 - 50$ kyr, 
as quoted above, we expect that there should be
between 0.07 -- 7 first cores in Perseus.  Thus, unless the very
longest lifetime estimates are correct, at least some (and possibly
all) are very young second cores (protostars).  Such objects would
also be consistent with the definition of a VeLLO, emphasizing that
these classes of objects are defined observationally and are not
necessarily mutually exclusive in terms of evolutionary stage.  

The number of ``starless'' cores that are truly starless is found by combining 
results for VeLLOs and candidate first cores.  Combined, 
\citet{2008ApJS..179..249D} and 
\citet{2012ApJ...745...18S} find that approximately 18\% -- 38\% of cores 
classified as starless prior to the launch of {\it Spitzer} in 2003 in fact
harbor low--luminosity sources, although the exact statistics are still 
quite uncertain.

\subsubsection{\textbf{PACS Bright Red Sources (PBRS)}}\label{sec_pbrs}

Using HOPS {\it Herschel} 70~\um\ imaging, 
\citet{2013ApJ...767...36S} searched for new protostars in Orion that were too 
faint at $\lambda \le 24~$~$\mu$m to be identified by \spitzer.  
They found 11 new objects with 70 \um\ and 160 \um\
emission 
that were either faint ($m(24) > 7$~mag) or undetected at 
$\lambda \le 24~$~$\mu$m.  In addition,
they found seven previously \spitzer\ identified protostars with
equally red colors 
(log~$\lambda F_{\lambda}(70)/\lambda F_{\lambda}(24) > 1.65$).  
These 18 PBRS are the reddest known protostars in
Orion A and Orion B. 
Fig.~\ref{fig:stutz_01} shows five representative PBRS.  

Although the emission at $\lambda \le 24$~$\mu$m is faint for all
PBRS, some are detected at 3.6--8.0 \um\ with {\it Spitzer}. 
These detections may
be due to scattered light and shocked gas in outflow cavities,
indicating the presence of outflows.  A total of eight PBRS
are undetected by \spitzer\ at 24~\um, leading to a lower limit for
their $\lambda F_{\lambda}(70)/\lambda F_{\lambda}(24)$ color. At
$\lambda \geq 70$~\um\ all PBRS have SEDs that are well characterized
by modified black--bodies (see Fig.~\ref{f.sed} for a
representative PBRS SED). The mean 70~\um\ flux of the PBRS is
similar to that of the rest of the HOPS sample of protostars, with a
comparable but somewhat smaller spread in values.

The overall fraction of PBRS to protostars in Orion is
$\sim$5\%. Assuming PBRS represent a distinct phase in protostellar
evolution (see further discussion of this assumption in \S 
\ref{sec_synthesis}), and further assuming a constant star formation rate and a 
protostellar duration of $\sim$0.5 Myr (\S \ref{sec_surveys}), the
implied duration of the PBRS phase is $\sim$25 kyr, averaging over
all Orion regions.  However, the spatial distribution of the PBRS
displays a striking non--uniformity compared to the distribution of
normal HOPS protostars: only 1\% of the protostars in Orion A are
PBRS, compared to $\sim$17\% in Orion B.  Whether this large variation
in the spatial distribution indicates a recent burst of star formation
or environmental differences remains to be determined.

Basic parameters of interest include \lbol, \tbol, and \lsl, as well
as modified black--body fits to the long--wavelength SEDs.  The PBRS
\lbol\ and \tbol\ distributions are shown in Fig.~\ref{f.blt};
\lbol\ spans a typical range compared to other protostars but
\tbol\ is restricted to very low values (\tbol\ $\leq 44$ K).
Similarly, \lsl\ values for PBRS occupy the extreme high end of
the distribution for all protostars in Orion
\citep[][see Fig.~\ref{fig_classification}]{2013ApJ...767...36S}.  
Modified black--body fits to the thermal portions of their SEDS yield PBRS 
envelope mass estimates in the range of $\sim
0.2$ --- $1.0$~\msun.  Radiative transfer models confirm that the
70~\um\ detections are inconsistent with externally heated starless
cores and instead require the presence of compact internal objects.
While the overall fraction of PBRS relative to protostars in Orion is
small ($\sim$5\%, see above), they are significant for their high
envelope densities.  Indeed, the PBRS have $70/24$ colors consistent
with very high envelope densities, near the expected Class 0/I
division or higher.  The above evidence points toward extreme youth,
making PBRS one of the few observational constraints we have on the
earliest phases of protostellar evolution, all at a common distance
with a striking spatial distribution within Orion.

\subsection{Synthesis of Observations}\label{sec_synthesis}

With the discoveries of VeLLOs, candidate FHSC, and PBRS, 
these large {\it Spitzer} and {\it Herschel} surveys of star-forming regions 
have expanded protostellar populations to objects that are both 
less luminous and more deeply embedded than previously known.  
These three object types are defined observationally, often based on
serendipitous discoveries, and are not necessarily mutually exclusive in terms 
of the physical stages of their constituents.  All
are likely heterogeneous samples encompassing true first
cores and protostars of varying degrees of youth.  Indeed, 
some of the lowest luminosity PBRS may in fact be first cores, and
others with low luminosities are consistent with the definition of
VeLLOs.  Furthermore, at least some, and possibly all, of the
objects identified as candidate first cores may not be true first cores but 
instead young protostars that have already evolved beyond the end of the 
first core stage.  Such objects would be consistent with the definition of 
VeLLOs since they all have low luminosities, and some may be detected 
as PBRS once {\it Herschel} observations of the Gould
Belt clouds are published.  Integrating these objects into the broader picture 
of protostellar evolution is of great importance and remains a subject of 
on-going study.  Further progress in characterizing the evolutionary status 
of these objects and using them as probes of the earliest stages of star 
formation depend on specific theoretical predictions for distinguishing 
between first cores and very young protostars followed by observations that 
test such predictions.

\section{\textbf{PROTOSTELLAR ACCRETION BURSTS AND VARIABILITY}}\label{sec_variability}

There is a growing body of evidence 
that the protostellar accretion process is variable and 
punctuated by short bursts of very rapid accretion (the 
``episodic accretion'' paradigm).  Indeed, optical variability was one of 
the original, defining characteristics of a young stellar object 
\citep{1945ApJ...102..168J,1952JRASC..46..222H}.  
We provide here a summary of the 
observational evidence for episodic mass accretion in the protostellar stage.

%

\subsection{Variability, Bursts, and Flares in Protostars}

There are two general classes of outbursting young stars known, 
FU Orionis type objects (FUors) and EX Lupi type objects (EXors), which 
differ in their outburst amplitudes and timescales (see accompanying chapter 
by {\it Audard et al}).  
As an example of an outbursting source, 
Fig.~\ref{f.sed} shows pre- and post-outburst SEDs of V2775 
Ori, an outbursting protostar in the HOPS survey area 
\citep{2012ApJ...756...99F}.  
Based on the current number of known outbursting young stars, and assuming 
that all protostars undergo repeated bursts, 
\citet{2011ApJ...736...53O} estimate that approximately 25\% of the total mass 
accreted during the protostellar stage does so during such bursts.  Some of 
the outbursting young stars detected to date are clearly still in the 
protostellar stage of evolution 
\citep[e.g.,][also see the accompanying chapter by {\it Audard et al.}]{2012ApJ...756...99F,2013arXiv1306.0666G}, but whether or not all protostars undergo 
large-amplitude accretion bursts remains an open question.


Infrared monitoring campaigns offer the best hope for answering this question 
\citep[e.g.,][]{2013ApJ...765..133J}, although we emphasize that not all 
detected variability is due to accretion changes; detailed modeling is 
necessary to fully constrain the variability mechanisms 
\citep[e.g.,][]{2012ApJ...748...71F,2013AJ....145...66F}.
\citet{2001AJ....121.3160C} and \citet{2002AJ....124.1001C} studied the 
near-infrared variability of objects in the Orion A and Chamaeleon I molecular 
clouds and identified numerous variable stars in each.  
\citet{2011ApJ...733...50M} 
presented preliminary results from YSOVAR, a {\it Spitzer} warm mission 
program to monitor young clusters in the mid-infrared, in which they identified 
over 100 variable protostars.  Once published, the final results of the 
program should provide robust statistics on protostellar variability 
\citep[][{\it Rebull et al.,} 2013, in prep.]{2011ApJ...733...50M}.
\citet{2013arXiv1306.2326W} found a variability fraction of 84\% among YSOs 
in Cygnus OB7 with two-year near-infrared monitoring.  
\citet{2012AJ....144..192M} found that 50\% of the YSOs in L1641 in Orion 
exhibited mid-infrared variability, with a higher fraction among the protostars 
alone compared to all YSOs, in multi-epoch {\it Spitzer} data spanning $\sim$ 
6 months.
\citet{2012ApJ...753L..35B} presented {\it Herschel} far-infrared monitoring 
of 17 protostars in Orion and found that 8 (40\%) show $>$10\% variability 
on timescales from 10 -- 50 days, which they attributed to accretion 
variability.  Finally, \citet{2013ApJ...768...93F} report results from the 
Palomar Transient Factory optical monitoring of the North American and Pelican 
Nebulae.  By monitoring at optical wavelengths they are very incomplete 
to the protostars in these regions, but all 
three of the protostars they detect show variability.

Another approach is to compare observations from different telescopes, which 
provides longer time baselines but introduces uncertainties from differing 
bandpasses and instrument resolutions.  
Three such studies have recently been published, although none had sufficient 
statistics to restrict their analysis to only protostars.  
\citet{2012ApJS..201...11K} compared {\it ISO} and {\it Spitzer} spectra for 
51 young stars and found that 79\% show variability of at least 0.1 
magnitudes, and 43\% show variability of at least 0.3 magnitudes, with all of 
the variability existing on timescales of about a year or shorter.  
\citet{2012MNRAS.420.1495S} compared 2MASS and UKIDDS near-infrared photometry 
with a time baseline of 8 years for 600 young stars and found that 50\% show 
$>$2$\sigma$ variability and 3\% show $>$0.5 magnitude variability, with the 
largest amplitudes seen in the youngest star-forming regions.  Based on their 
statistics they derived an interval of at least 2000 -- 2500 years between 
successive bursts.
\citet{2013MNRAS.430.2910S} compared {\it Spitzer} and {\it WISE} photometry 
with 5 year time baselines for 4000 young stars and identified 1 -- 4 strong 
burst candidates with $>$1 magnitude increases between the two epochs.  Based 
on these statistics they calculated a typical interval of 10,000 years between 
bursts.  

At present, neither direct monitoring campaigns nor comparison of data from 
different telecopes at different epochs offer definitive statistics on 
protostellar variability or the role variability plays in shaping the 
protostellar luminosity distribution.
However, they do demonstrate that variability is common among all YSOs, and 
protostars in particular, and they do offer some of the first statistical 
constraints on variability and burst statistics.  We anticipate continued 
progress in this field in the coming years.

\subsection{Accretion Variability Traced by Outflows}

Molecular outflows are driven by accretion onto protostars and are 
ubiquitous in the star formation process (see accompanying chapter by 
{\it Frank et al.}).  Any variability in the underlying accretion process 
should directly correlate with variability in the ejection process.  Indeed, 
many outflows show clumpy structure that can be interpreted as arising from 
separate ejection events.  Although such clumpy structures can also arise 
from the interaction between outflowing gas and a turbulent medium even 
in cases where the underlying ejection process is smooth 
\citep{2011ApJ...743...91O}, many outflows with such structure also display 
kinematic evidence of variability.  In particular, outflow clumps in 
molecular CO gas, which are often spatially coincident with near-infrared 
H$_2$ emission knots, are found at higher velocities than the rest of the 
outflowing gas, and within these clumps the velocities follow Hubble laws 
(increasing velocity with increasing distance from the protostar).  This 
creates distinct structures in position-velocity diagrams called 
``Hubble wedges'' by 
\citet{2001ApJ...554..132A}, which are consistent with theoretical expectations 
for prompt entrainment of molecular gas by an episodic jet 
\citep[e.g.,][and references therein]{2001ApJ...554..132A}.  
One of the most recent examples of this phenomenon is presented by 
\citet{2013arXiv1304.0674A} 
for the HH46/47 outflow using ALMA data and is shown in Fig.~\ref{fig_hh46}.  
The timescales of the 
episodicity inferred from the clump spacings and velocities range from 
less than 100 years to greater than 1000 years 
\citep[e.g., ][]{1991A&A...251..639B,2009ApJ...699.1584L,2013arXiv1304.0674A}.

\begin{figure}[t]
\epsscale{1.0}
\plotone{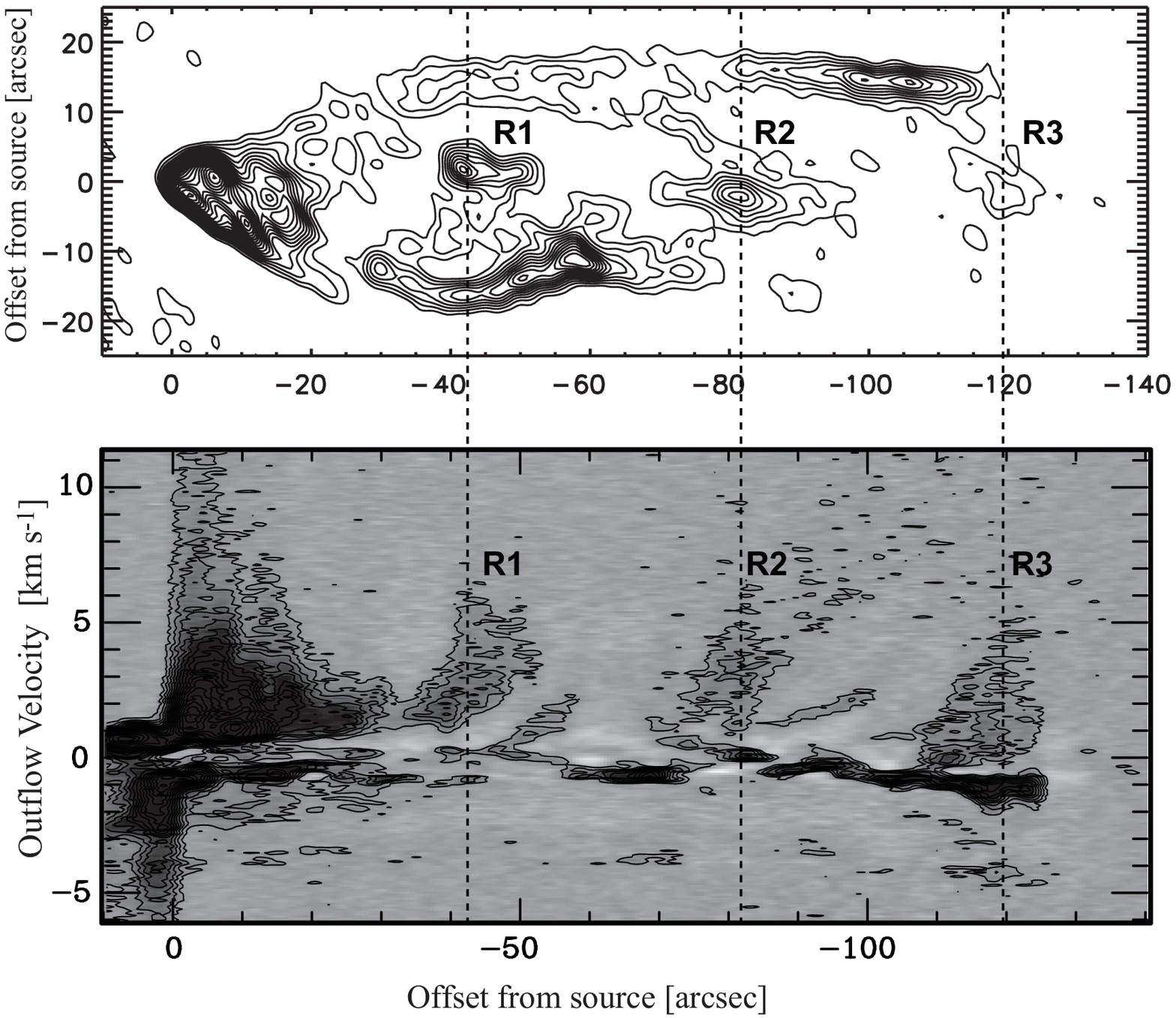}
\caption{\label{fig_hh46}\small 
{\it Top panel:}  Integrated \cojone\ emission of the redshifted lobe of 
the outflow driven by the HH46/47 protostellar system based on the data 
presented by \citet{2013arXiv1304.0674A}, with the positions of three 
clump-like structures labeled.
{\it Bottom panel:}  Position-velocity diagram along the axis 
of the redshifted lobe of the same outflow.  The three clumps clearly 
exhibit higher velocities than the rest of the outflowing gas, with each 
clump exhibiting its own Hubble law.}
 \end{figure}

Additional evidence for accretion variability comes from the integrated 
properties of molecular outflows.  In a few cases, low luminosity protostars 
drive strong outflows implying higher {\it time-averaged} accretion rates 
traced by the integrated outflows than the {\it current} rates traced by the 
protostellar luminosities 
\citep{2006ApJ...651..945D,2010ApJ...721..995D,2010ApJ...709L..74L,2012AJ....144..115S}.  
As discussed by 
\citet{2010ApJ...721..995D}, the amount by which the accretion rates must 
have decreased over the lifetimes of the outflows are too large to 
be explained solely by the slowly declining accretion rates predicted by 
theories lacking short-timescale variability and bursts 
(see \S \ref{sec_accretion}).

\subsection{Chemical Signatures of Variable Accretion}


If the accretion onto the central star is episodic, driving substantial
changes in luminosity, there can be observable effects on the chemistry
of the infalling envelope and these can provide clues to the luminosity
history.  Several authors have recently explored these effects using 
chemical evolution models, as discussed in more detail in the accompanying 
chapter on episodic accretion by {\it Audard et al.}  
A particular opportunity to trace the luminosity
history exists in the absorption spectrum of \cotwo\ ice; observations of the 
15.2 \um\ \cotwo\ ice absorption feature toward low-luminosity protostars with 
{\it Spitzer} spectroscopy have provided strong evidence for past 
accretion and luminosity bursts in these objects 
\citep[][see \S \ref{sec_infalling}]{2012ApJ...758...38K}.


\section{\textbf{THE FORMATION AND EVOLUTION OF PROTOSTELLAR DISKS}}\label{sec_disks}

\subsection{Theoretical Overview}

The formation of a circumstellar disk is a key step in the formation of planets and/or binary
star systems, where in this chapter ``disk'' refers to a rotationally 
supported, Keplerian structure. 
Disks must form readily during the star formation process
as evidenced by their ubiquity during the T-Tauri phase \citep[see Fig.~14 of][]{2007ApJ...662.1067H}. In order to form a disk during the collapse
phase, the infalling material must have some specific angular momentum \citep{1981Icar...48..353C};
disks may initially start small and grow with time. The
formation of the disk is not dependent on whether the angular momentum derives 
from initial cloud rotation \citep{1981Icar...48..353C,1984ApJ...286..529T} or the turbulent medium 
in which the core formed \citep[e.g.][]{2010ApJ...725.1485O}, but the subsequent growth of the disk will depend on the 
distribution of angular momentum.  After disk formation, 
most accreting material must be processed through the disk. Moreover, models for accretion bursts
and outflows depend on the disk playing a dominant role (see \S \ref{sec_accretion} and the accompanying chapters by {\it Li et al.}, {\it Audard et al}, and 
{\it Frank et al.}).  
Finally, disk rotation, when detectable, allows the only means of direct determination of protostellar masses (see the accompanying chapter by {\it Dutrey et al.}), since spectral types are unavailable for most protostars that are too deeply embedded to detect with optical and/or near-infrared spectroscopy, and mass is only one of several parameters that determines the luminosity of a protostar (see \S \ref{sec_lum}).


Numerical hydrodynamic models of collapsing, rotating clouds
predict the formation of massive and extended disks in the early embedded stages
of protostellar evolution \citep{2009ApJ...692.1609V}, with the disk mass increasing 
with the protostellar mass \citep{2011ApJ...729..146V}. 
However, these and earlier studies neglected the role of magnetic fields.  
\citet{2003ApJ...599..363A} presented ideal magneto-hydrodynamic (MHD) 
simulations examining the effects of magnetic braking and showed that 
collapsing material drags the magnetic field inward and increases the field 
strengths toward smaller radii.
The magnetic field is anchored to the larger-scale envelope and 
molecular cloud, enabling angular momentum to be removed from the collapsing inner
envelope, preventing the formation of a rotationally supported disk.

This finding is referred to as the ``magnetic braking catastrophe'' 
and had been verified in the ideal MHD limit analytically by \citet{2006ApJ...647..374G}
and in further numerical simulations by \citet{2008ApJ...681.1356M}.
These studies concluded that the magnetic braking efficiency must
be reduced in order to enable the formation of rotationally supported disks. 
Non-ideal MHD simulations have shown that Ohmic dissipation can enable the 
formation of only very small disks (R $\sim$10 \rsun) that are not 
expected to grow to hundred-AU scales until after most of the envelope has 
dissipated in the Class II phase 
\citep{2010A&A...521L..56D,2011PASJ...63..555M}.  
Disks with greater radial extent and larger masses can form if magnetic fields 
are both relatively weak and misaligned relative to the rotation axis 
\citep{2012A&A...543A.128J}, 
and indeed observational evidence for such misalignments has been claimed by 
\citet{2013ApJ...768..159H}.  
\citet{2013ApJ...767L..11K} estimated that 10\% - 50\% of protostars might 
have large (R $>$ 100 AU), rotationally supported disks
by applying current observational constraints on the strength and alignment 
of magnetic fields to the simulations of \citet{2012A&A...543A.128J}.

\subsection{Observations of Protostellar Disks}

The earliest evidence for the existence of protostellar disks comes from models 
fit to the unresolved infrared and (sub)millimeter continuum SEDs of protostars 
\citep[e.g.,][]{1986ApJ...308..836A,1993ApJ...414..676K,1994ApJ...420..326B,1997ApJ...481..912C,2003ApJ...598.1079W,2006ApJS..167..256R,2004ApJS..154..396Y}.  
Models including circumstellar disks provided good fits to the observed SEDs, 
and in many cases disks were required in order to obtain satisfactory fits.  
However, SED fitting is often highly degenerate between disk and inner core 
structure, and even in the best cases does not provide strong constraints 
on disk sizes and masses, the two most relevant quantities for evaluating 
the significance of magnetic braking.  Thus in the following sections we 
concentrate on observations that do provide constraints on these quantities, 
with a particular focus on (sub)millimeter interferometric observations.

\subsubsection{Class I Disks}

Due to complications with separating disk and envelope emission, 
a characterization of Class I disks on par with Class II disks \citep[e.g.][]{2010ApJ...723.1241A} has not yet been possible.
Several approaches have been taken toward characterizing Class I disks, including focusing on SEDs, scattered light imaging, and millimeter emission. 

Analysis of millimeter emission requires invoking models to disentangle 
disk and envelope emission.  This approach was carried out by 
\citet{2009A&A...507..861J}, who found 10 Class I disks with 
$M \sim$ 0.01 \msun, although the disks were not resolved.
\citet{2012ApJ...755...23E} modeled the disk and envelope properties
of eight Class I protostars in Taurus combining the SEDs, millimeter imaging, and near-infrared 
scattered light imaging \citep[see also][]{2005ApJ...635..396E}. 
The median radius of the sample was found to be 250 AU and the median
disk mass was 0.01 $M_{\sun}$, but the derived parameters for each source 
depended strongly on the weighting of the model fits.  They also did not 
resolve the disks, with only $\sim$1\arcsec\ resolution.

The most definitive constraints on Class I disks have come from the detailed modeling of edge-on disks resolved
in both near-infrared scattered light and millimeter dust emission.  Two 
recent examples of such work include studies of the edge-on Class I protostars 
IRAS 04302+2247 and CB26.  IRAS 04302+2247 was found to have a disk with 
$R \sim$ 300 AU and $M \sim$ 0.07 \msun\ 
\citep{2008ApJ...674L.101W,2013A&A...553A..69G}.  CB26 has a disk with 
$R \sim$ 200 AU and a possible inner hole of 45 AU cleared by an undetected 
binary companion \citep{2009A&A...505.1167S}.  These representative cases 
illustrate that relatively large, massive disks may be common in the 
Class I stage.

\begin{figure*}
\includegraphics[angle=-90, scale=0.66]{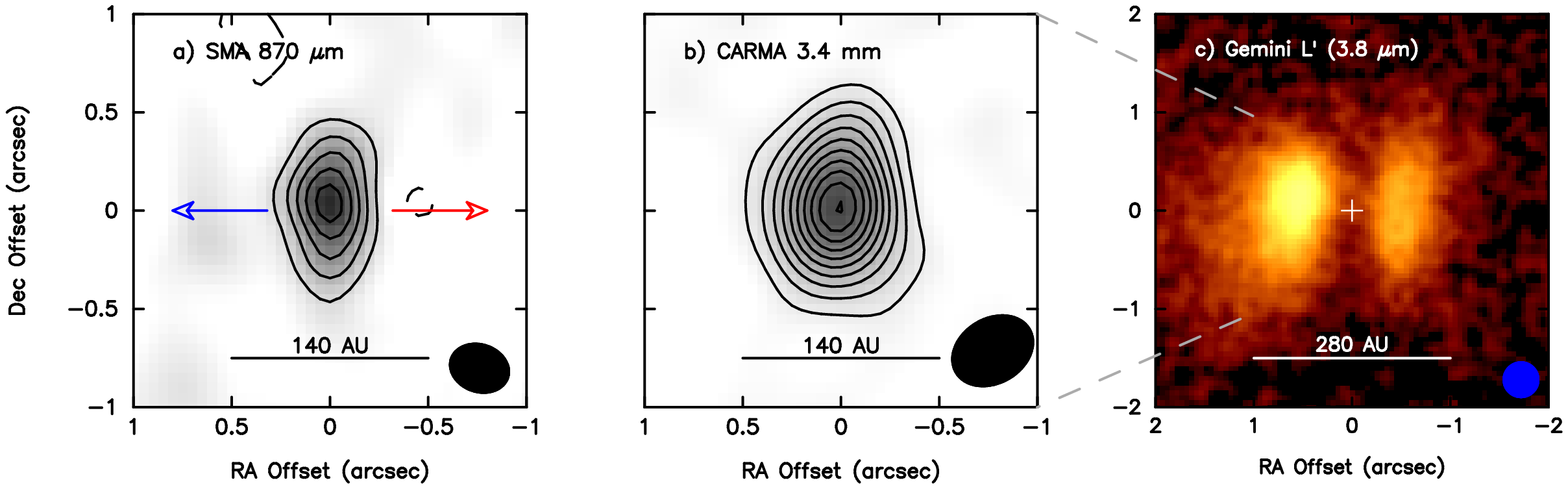}
\caption{\small Images of the edge-on disk around the protostar L1527 from 
\citep{2012Natur.492...83T}.  
High-resolution images of L1527 showing the disk in dust continuum emission and scattered light are shown at wavelengths of 870 \micron\ from the SMA
(a), 3.4 mm from CARMA (b), and 3.8 \micron\ from Gemini (c). 
The Gemini image is shown on a larger scale.  
The sub/millimeter images are elongated in the direction 
of the dark lane shown in panel (c), consistent with
an edge-on disk in this Class 0 protostellar system. 
The outflow direction is indicated by arrows in panel (a).  
The cross in panel (c) marks the central position of the disk
from the SMA images. 
}
\label{L1527-disk}
\end{figure*}

Sensitive molecular line observations are also enabling the kinematic structure 
of Class I disks to be examined. A number of Class I disks have been found to be rotationally-supported 
\citep[e.g.][]{2009A&A...507..861J,2007A&A...461.1037B,2008A&A...481..141L,2012ApJ...754...52T, 2013arXiv1305.2668H}.
Thus far, the number of Class I protostellar mass measurements is only 6,
and their masses range between 0.37 and 2.5 $M_{\sun}$. The disk masses are typically less than 10\% of
the stellar masses, modulo uncertainties in calculating mass from dust emission and modeling systematics.
\citet{2009A&A...507..861J} pointed out possible discrepancies between models of disk formation from the
simple picture of an infalling, rotating envelope \citep{1984ApJ...286..529T}. Semi-analytic
 models of protostellar collapse from \citet{2009A&A...495..881V} and hydrodynamic models 
of \citet{2009ApJ...692.1609V,2011ApJ...729..146V} 
seem to suggest much more massive disks (up to 40\%--60\% that of the star) 
than revealed by observations.

\subsubsection{Class 0 Disks}

Observations of Class 0 disks face challenges similar to the Class I disks, 
but with the added complication of more massive envelopes responsible for $\sim$90\% of the emission at millimeter wavelengths \citep{2000ApJ...529..477L}. 
The SEDs of Class 0 protostars are dominated by the far-infrared component
and the near-infrared is generally dominated by scattered light from the 
outflow cavity.  Much like the Class I systems, it is impossible to garner any 
constraints of the disk properties in Class 0 systems from the SEDs alone. 

B335 was one of the first Class 0 systems to be examined in great
detail and found to have a disk with $R <$ 60 AU and $M \sim$ 0.0014 \msun\ 
\citep{2003ApJ...596..383H}; while that study assumed a distance of 250 pc, 
we have scaled these numbers to the more recent estimates that 
place B335 at $\sim$150 pc \citep{2008ApJ...687..389S}.  Among other Class 0 protostars, HH211 shows 
extended structure perpendicular to the outflow that may indicate a 
circumstellar disk \citep{2009ApJ...699.1584L}.  
On the other hand, the Class 0 protostar L1157-mm shows no evidence of 
resolved disk structure on any scale with resolutions as fine as 0\farcs3, 
implying any disk present must have $R <$ 40 AU and $M <$ 0.004--0.025 \msun\ 
\citep{2012ApJ...756..168C}, and ALMA observations of IRAS 16293-2422 
revealed only a very small disk with $R \sim 20$ AU.
\citep{2013ApJ...768..110C, 2013ApJ...764L..14Z}.  
\citet{2007ApJ...659..479J} and \citet{2009A&A...507..861J} coupled a 
submillimeter survey of 10 Class 0 protostars 
with models for envelope emission and found Class 0 disks with $M \sim$ 0.01 
\msun\ (with significant scatter), and no evidence of disk mass growth between 
Class 0 and I sources.  
On the other hand, \citet{2010A&A...512A..40M} did not find any evidence for 
disk structures $>$ 100 AU in a millimeter survey of 5 Class 0 protostars 
at sub-arcsecond resolution (linear resolution $<$100 AU).  

The clearest evidence of Class 0 disks thus far is found toward the protostars 
L1527 IRS and VLA1623A.  For L1527, 
high resolution scattered light and (sub)millimeter imaging observations found 
strong evidence of an edge-on disk, \citep[][see Fig.~\ref{L1527-disk}]{2012Natur.492...83T}. Furthermore, molecular line observations were found to trace 
Keplerian rotation, implying a protostellar mass of 0.19 $\pm$ 0.04 \msun. 
The mass of the surrounding envelope is $\sim$5 times larger than the 
protostar, and the disk has $R = 70-125$ AU and $M \sim 0.007$ \msun\ 
\citep{2012Natur.492...83T,2013arXiv1305.3604T}.
L1527 IRS is classified as Class 0 based on its SED.  While its edge-on nature 
can bias this classification \S \ref{sec_surveys}, it does have 
a more massive and extended envelope than a typical Class I 
protostar in Taurus and is consistent with the Stage 0 definition of 
$M_{*}$ $<$ $M_{env}$.  
For VLA1623A, \citet{2013ApJ...764L..15M} and \citet{2013arXiv1310.8481M} 
identified a confirmed disk via detection of Keplerian rotation, with 
$R \sim 150$ and $M \sim 0.02$ \msun.

\subsection{Summary of Protostellar Disks}

Current observations clearly indicate that there are large (R $>$ 100 AU) disks present in 
the Class I phase but their properties have been difficult to detail en masse 
thus far. There is evidence for large disks in \textit{some} Class 0 systems 
(L1527 IRS, VLA 1623, HH211), but more observations with sufficient
resolution to probe scales where disks dominate the emission are needed for a 
broader characterization. Efforts to detect more disk structures in the nearby 
Perseus molecular cloud are underway and have found two strong candidates for 
Class 0 disks ({\it Tobin et al.}, in prep).

Despite remarkable progress in characterizing disks around protostars in the 
last decade, several puzzles remain.  First, if disks form via a process 
similar to the rotating collapse model, the observed disk masses are about 
an order of magnitude lower than theoretical estimates for a given 
protostellar mass 
\citep{2009ApJ...692.1609V,2009A&A...495..881V,1999ApJ...525..330Y}. Second, 
models considering magnetic braking seem to suggest small disks until the 
Class II phase, unless there are misaligned fields; however, observations seem 
to indicate at least some 
large disks in the Class 0 and I phases. Lastly, it is not clear 
if there is any observational evidence for magnetic braking during collapse, 
as \citet{2013arXiv1305.6877Y} recently showed that angular momentum is 
conserved in two Class 0 systems.

Moving forward, ALMA will provide key observations for determining the 
presence and properties of Class 0 disks, offering significant improvements 
over the resolution and sensitivity limits of current interferometers.  
By detecting Keplerian rotation, ALMA holds the promise to enable mass 
measurements for large numbers of protostars, possibly even directly observing 
for the first time the protostellar mass function \citep{2010ApJ...716..167M}.  
At the same time, the upgraded Very Large Array (VLA)
 has very high sensitivity to dust 
continuum emission at 7 mm and may also contribute to the characterization of 
protostellar disks.

\section{\textbf{THE EVOLUTION OF INFALLING MATERIAL}}\label{sec_infalling}

Silicate dust and ices are the building blocks of comets and, ultimately, Earth-like planets, and both undergo significant chemical and structural (crystallinity) changes during the protostellar stage.  Understanding how these materials evolve as they flow from molecular cores to protostellar envelopes and disks is an essential component in determining how matter is modified in planet-forming disks.  While most of this Chapter focuses on macroscopic protostellar evolution, we now examine the microscopic evolution of solid-state matter surrounding protostars.  

Mid-infrared spectra toward low-mass protostars contain a wealth of solid-state absorption features due to the vibrational modes of silicate dust and many molecular ice species \cite[e.g.,][]{2008ApJ...678..985B, 2008ApJS..176..184F}.  These materials are inherited from the dense interstellar medium, and are present in the infalling envelopes and accretion disks surrounding protostars. Within such environments, radiative and mechanical processing by UV photons and accretion shocks may sublimate or modify the properties of dust and ices.  Indeed, theoretical disk evolutionary models of low-mass protostars predict that infalling ices, with sublimation temperatures greater than \emph{T} $\approx$ 50 K, remain on grain surfaces after reaching the accretion disk \citep{2009A&A...495..881V}, suggesting that at least a fraction of cometary ices are of protostellar origin.  Furthermore, the formation of crystalline silicates are predicted to have occurred during the earliest phases of disk formation, when a fraction of the amorphous silicate dust falls close to the central protostar and is subsequently heated to temperatures of \emph{T} = 800-1200 K \citep{2006ApJ...640L..67D, 2010A&A...519A..28V}.  

While it is generally assumed that solids present in the infallling envelopes of low-mass protostars are mostly pristine without significant processing, the {\it Spitzer Space Telescope} has now revealed that such matter shows clear evidence for thermal processing.  In this section, we review the current observational evidence for high- to low-temperature processing of silicates and CO$_{2}$ ice, and their implications for both protostellar evolution and the inventory of matter delivered to planet-forming disks.

\subsection{Thermal Processing of Amorphous Silicates}

\begin{figure}[t]
\centering
\includegraphics[width=0.488\textwidth]{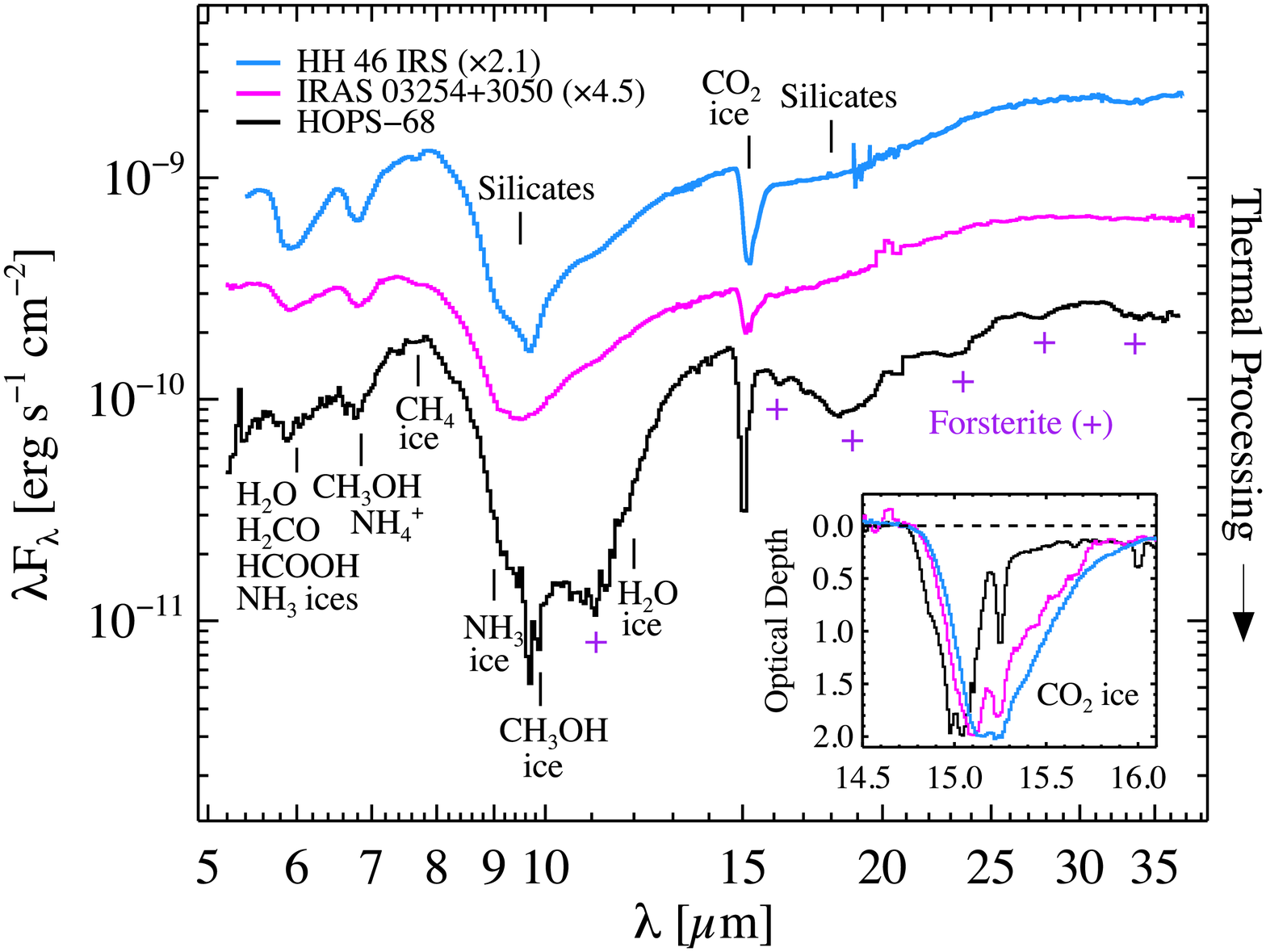}
\caption{\small \emph{Spitzer}-IRS spectra of low-mass prototstars, ordered by increasing evidence for thermal processing, with solid-state features indicated.  Unlike HOPS 68, the spectra of HH 46 IRS and IRAS 03254+3050 show no signs for the presence of crystalline silicates (forsterite).  Insert: arbitrarily scaled CO$_{2}$ ice optical depth spectra of protostars, exhibiting various degrees of ice processing that is evident by their double-peaked substructure.  Data are presented in \citet{2008ApJ...678..985B, 2008ApJ...678.1005P, 2011ApJ...733L..32P, 2013ApJ...766..117P}.}
\label{fig:spectra}
\end{figure}

Prior to the launch of the {\it Spitzer Space Telescope}, little knowledge of the silicate dust surrounding low-mass protostars existed.  Because observations with the {\it Infrared Space Observatory} were sensitivity-limited to the most luminous, massive protostars, detailed spectral modeling was possible for only a few high-mass protostars \citep{1999A&A...349..267D, 2000ESASP.456..183D}.  Similar to the silicate composition of the diffuse interstellar medium, which is thought to be almost entirely amorphous in structure \citep{2005ApJ...633..534K}, only a small fraction (1\%-2\%) of the silicate dust mass present in protostellar envelopes was predicted to be in crystalline form during the pre-{\it Spitzer} era.  

The high sensitivity of the Infrared Spectrograph (IRS) on board the {\it Spitzer Space Telescope} permitted the routine detection of crystalline silicates emission features toward T Tauri stars \cite[e.g.,][]{2009ApJS..182..477S, 2010A&A...520A..39O} and comets \cite[e.g.,][]{2007Icar..191S.223L, 2009P&SS...57.1133K}.  Crystalline silicate emission features are also detected toward the Serpens protostellar binary SVS 20 \citep{2005ApJ...629..897C}; however, a recent evaluation of its \emph{Spitzer}-IRS spectrum suggests that SVS 20 may be a T Tauri star possessing a flattened, settled disk \citep{2010ApJ...714..778O}.

In contrast, more than 150 low-mass protostars have been studied in the literature, and their {\it Spitzer}-IRS spectra exhibit 10 and 20 $\micron$ amorphous silicate absorption features that are almost always characterized by broad, smooth profiles, lacking any superimposed narrow structure.  However, the one exception is HOPS 68, a low-mass embedded protostar situated in the Orion Molecular Cloud complex \citep{2011ApJ...733L..32P}.  The {\it Spitzer}-IRS spectrum toward HOPS 68 exhibits narrow {\it absorption} features of forsterite 
(Fig.~\ref{fig:spectra}), indicating that a significant fraction ($\lesssim$17\%) of amorphous silicates within its infalling envelope have experienced strong thermal processing (\emph{T} $\gtrsim$ 1000 K).  Although the mechanisms responsible for such processing are still not fully understood, it is proposed that amorphous silicates were annealed or vaporized within the warm inner region of the disk and/or envelope and subsequently transported outward by entrainment in protostellar outflows.  Alternatively, an {\it in situ} formation by shock processing in the outflow working surface may be responsible for the production of crystalline silicates within the envelope of HOPS 68.   

The detection of crystalline silicates in cometary material necessitates a process for transporting thermally processed silicates to the cold outer regions of planet-forming disks \citep[e.g.,][]{2011ApJ...740....9C} or requires an {\it in situ} formation route to be present at large radial distances \citep[e.g.,][]{2011ApJ...728L..45V, 2011MNRAS.416L..50N}.  The detection of crystalline silicates toward HOPS 68 demonstrates that at least in one case, thermally processed silicates may be delivered to the outer accretion disk by infall from the protostellar envelope.  Of course, HOPS 68 may be a unique case.  However, on the basis of its peculiar SED, it is argued that HOPS 68 has a highly flattened envelope structure that provides a line of sight to the inner region without passing through the intervening cold, outer envelope \citep{2011ApJ...733L..32P, 2013ApJ...766..117P}.  Thus, HOPS 68 may provide a rare glimpse into the thermally processed inner regions of protostellar envelopes.

\subsection{Thermal Processing of Interstellar CO$_{2}$ Ice}

Interstellar ices serve as excellent tracers of the thermal history of protostellar environments \citep{2004ASPC..309..547B}.  As material falls from the outer envelope to the accretion disk, grains are subject to elevated temperatures while traversing the inner envelope region, resulting in the crystallization of hydrogen-rich ices or the sublimation of CO-rich ices \citep[e.g.,][]{2009A&A...495..881V}.  Moreover, protostellar feedback mechanisms, such as outflow-induced shocks and episodic accretion events, may also be responsible for the crystallization and sublimation of interstellar ices \citep{1998ApJ...499..777B, 2007JKAS...40...83L}.

CO$_{2}$ ice has proven to be a sensitive diagnostic of the thermal history toward protostars.  Pure CO$_{2}$ ice may be produced by (1) segregation of CO$_{2}$ from hydrogen-rich ice mixtures (e.g., H$_{2}$O:CH$_{3}$OH:CO$_{2}$ and H$_{2}$O:CO$_{2}$) or (2) thermal desorption of CO from a CO:CO$_{2}$ ice mixture at temperatures of \emph{T} = 30-60 K and \emph{T} = 20-30 K, respectively \citep{1998A&A...339L..17E, 2009A&A...505..183O, 2006A&A...451..723V}.  Spectroscopically, thermally processed mixtures of hydrogen- and CO-rich CO$_{2}$ ices produce a double-peaked substructure in the 15.2 $\micron$ CO$_{2}$ ice bending mode profile, characteristic of pure, crystalline CO$_{2}$ ice (Fig. \ref{fig:spectra}).

The presence of pure CO$_{2}$ ice is generally interpreted as a result of thermal processing of pristine icy grains in cold, quiescent regions of dense molecular clouds.  To date, $\sim$100 high-resolution {\it Spitzer}-IRS spectra toward low-mass protostars have been described in the literature, and the majority of the CO$_{2}$ is found in hydrogen- and CO-rich ices typical of molecular cloud cores \citep{2008ApJ...678.1005P, 2009ApJ...694..459Z, 2011ApJ...730..124C, 2012ApJ...758...38K}.  However, nearly 40\% of the spectra show some evidence for the presence of pure CO$_{2}$ ice, suggesting that low-mass protostars have thermally processed inner envelopes.  Among these, relatively large abundances ($\sim$15\% of the total CO$_{2}$ ice column density) of pure CO$_{2}$ ice have been detected toward a sample of low-luminosity (0.08 \emph{L}$_{\sun}$ $\lesssim$ \emph{L} $\lesssim$ 0.69 \emph{L}$_{\sun}$) embedded protostars \citep{2012ApJ...758...38K}.  Because the present environments of low-luminosity protostars do not provide the thermal conditions needed to produce pure CO$_{2}$ in their inner envelopes, a transient phase of higher luminosity must have existed some time in the past \citep{2012ApJ...758...38K}.  The presence and abundance of pure CO$_{2}$ ice toward these low-luminosity protostars may be explained by episodic accretion events, in which pure CO$_{2}$ is produced by distillation of a CO:CO$_{2}$ mixture during each high-luminosity transient phase.  Other accretion models have not yet been tested and may also explain the observed presence of pure CO$_{2}$ ice toward low-luminosity protostars; however, continuous accretion models with monotonically increasing luminosity cannot reproduce the observed abundances of CO$_{2}$ ice and C$^{18}$O gas \citep{2012ApJ...758...38K}.

The highest level of thermally processed CO$_{2}$ ice is found toward the moderately luminous (1.3 \emph{L}$_{\sun}$) protostar HOPS 68 \citep{2013ApJ...766..117P}.  Its CO$_{2}$ ice spectrum reveals an anomalous 15.2 $\micron$ bending mode profile that indicates little evidence for the presence of unprocessed ice.  Detailed profile analysis suggests that 87\%-92\% of its CO$_{2}$ ice is sequestered as spherical, CO$_{2}$-rich mantles, while typical interstellar ices are dominated by irregularly shaped, hydrogen-rich CO$_{2}$ mantles \citep[e.g.,][]{2008ApJ...678.1005P}.  The nearly complete absence of unprocessed ices along the line of sight to HOPS 68 is best explained by a highly flattened envelope structure, which lacks cold absorbing material in its outer envelope, and possesses a large fraction of material within its inner (10 AU) envelope region.  Moreover, it is proposed that the spherical, CO$_{2}$-rich ice mantles formed as volatiles rapidly froze out in dense gas, following an energetic but transient event that sublimated primordial ices within the inner envelope region of HOPS 68.  The mechanism responsible for the sublimation is proposed to be either an episodic accretion event or shocks in the interaction region between the protostellar outflow and inner envelope.  Presently, it is unknown if such feedback mechanisms are also responsible for the presence of crystalline silicates toward HOPS 68.

\section{\textbf{DOES ENVIRONMENT MATTER?}}\label{sec_environment}

Low mass stars form in a wide range of environments: isolated globules, small groups, and rich clusters of low and high mass stars \citep[e.g.][]{2009ApJS..184...18G,2013A&A...551A..98L}, with no apparent dichotomy between clustered and isolated star formation \citep{2010MNRAS.409L..54B}.  This range of environments provides a natural laboratory for studying low mass star formation in different physical conditions, with the goal of elucidating how those physical conditions may guide the formation of the stars.  In this section, we explore how the environment affects the incidence of protostars and their properties. We define local environment as the region outside the immediate core/envelope, i.e. beyond 10,000-20,000~AU \citep{2008ApJ...684.1240E}.  Thus, the environmental conditions would include the properties of the gas in the surrounding molecular cloud or the density of young stars in the vicinity of the protostar, but would exclude the local molecular core/envelope as well as companions in a multiple star system. 

Many of the environmental conditions have yet to be measured over the spatial extent of the {\it Spitzer} and {\it Herschel} surveys.  Thus we focus on three measures of the environment: the gas column density, cloud geometry, and the surface density of YSOs, to address four specific questions.  

\subsection{Does the Incidence and Density of Protostars Depend on the Local Gas Column Density?}
\label{sec:incidence}

A combination of (sub)millimeter surveys for cores and extinction maps constructed from 2MASS and {\it Spitzer} photometry of background stars have enabled an examination of the incidence of molecular cores as a function of the column density of gas smoothed over spatial scales of 0.2 to 1~pc (see also the accompanying chapter by {\it Padoan et al.}~for a similar discussion in the context of star formation laws).  Millimeter continuum surveys of the Ophiuchus and Serpens clouds show that the cores  are found in environments where the column densities exceed $A_V = 7$~mag  \citep{2004ApJ...611L..45J,2008ApJ...684.1240E}, consistent with the model of photoionization regulated star formation in magnetized clouds \citep{1989ApJ...345..782M}.    However, in the Perseus cloud, \citet{2008ApJ...684.1240E} found that 25\% of cores are located at $A_V < 7$~mag \citep[however, see][]{2006ApJ...646.1009K}, and \citet{2005A&A...440..151H} showed that the probability of finding a sub-millimeter core in the Perseus cloud increased continuously with the integrated intensity of the C$^{18}$O ($1 \rightarrow 0$) line to the 3rd power.   In total, these results indicate a rapid rise in the incidence of cores with column density, but with a small number of cores detected below $A_V = 7$~mag. 
  
A similar result comes from the examination of {\it Spitzer} identified infrared protostars. In a combined sample of seven molecular clouds, \citet{2011ApJ...739...84G} and \citet{2012ApJ...752..127M} showed that the surface density of dusty YSOs (i.e. Class 0/I {\it and} Class II objects) increases with the 2nd to 3rd power of the gas column density.  \citet{2010ApJ...723.1019H} also found a steep rise in the surface density of c2d and GB protostars with the column density of gas.  Although {\it Spitzer} identified candidate protostars at column densities as low as 60~M$_{\odot}$ pc$^{-2}$ \citep[$A_V = 3$ mag, ][]{2011ApJ...739...84G}, the protostars are preferentially found in much denser environments, with a rapid, power-law like rise in the density of protostars with increasing gas column density.   Currently, it is not clear whether the rapid rise in the incidence of cores and density of protostars with column density is due to the inhibition of star formation in regions of low column density, as argued by the photoionization regulated star formation model, or a non-linear dependence of the density of protostars on the gas column density, such as that predicted by Jeans fragmentation \citep{1985MNRAS.214..379L}.

\subsection{What is the Connection between Cloud Geometry and the Spatial Distribution of Protostars?}

The elongated and filamentary nature of interstellar clouds has been evident since the earliest optical observations of dark clouds \citep{1907ApJ....25..218B}.  With improvements in sensitivity and resolution, it has become clear that star-forming molecular clouds are complex filamentary networks, with filamentary structure on scales from hundreds of pc to hundreds of AU \citep{1962ApJS....7....1L,1979ApJS...41...87S,2009ApJ...700.1609M,2010ApJ...712.1010T,2010A&A...518L.102A,2010A&A...518L.100M}. Large-scale filaments harbor star-forming cores and protostars, especially at bends, branch points, or ÒhubsÓ which host groups or clusters of cores and protostars \citep{2009ApJ...700.1609M,2012A&A...540L..11S}.  Filaments may both accrete gas and channel its flow along their length \citep{2011ApJ...740L...5C}; recent observations of one young cluster found that flow onto and along filaments may feed gas to the central, cluster-forming hub \citep{2013ApJ...766..115K}.  The filamentary structure of star-forming gas in Orion and the corresponding distribution of protostars is apparent in Fig.~\ref{fig:hops_comp}.


Self-gravitating filaments may fragment with a characteristic spacing \citep{1985MNRAS.214..379L,1992ApJ...388..392I,1997ApJ...480..681I}. Indeed, the spacings determined from the  initial temperature, surface density, and length of observed filaments predict the approximate number of low-mass stars in the Taurus complex and other complexes which lack rich clusters  \citep{2002ApJ...578..914H,2011ApJ...735...82M}.  A relationship between the spacing of protostars with gas column density is clearly apparent in Fig.~\ref{fig:hops_comp}, which shows recent HOPS observations of two filamentary regions within the Orion cloud.  The OMC-2/3 region is directly north of the Orion Nebula and is considered a northern extension of the ONC \citep{2008hsf1.book..590P,2012AJ....144..192M}.  In contrast, L1641S is a more quiescent region in the southern part of the L1641 cloud \citep{2000AJ....120.3139C,2008hsf1.book..621A}. Although both regions host multi-parsec filaments with mass to length ratios exceeding the limit for a stable, thermally supported filament \citep{2010A&A...518L.102A,2012A&A...542A..77F}, the average column density calculated above a cutoff of $N(H_2) = 3 \times 10^{21}$~cm$^{-2}$ is twice as high in OMC2/3: $N(H_2) = 1.4 \times 10^{22}$ cm$^{-2}$ in OMC2/3 as compared to $ 7.6 \times 10^{21}$~cm$^{-2}$ in L1641S.   The density of protostars is correspondingly higher in the OMC2/3 region; with the typical protostellar spacing being 6500 AU and 14000 AU in OMC2/3 and L1641, respectively.   


\begin{figure}[lineheight,t]
\includegraphics[trim=0cm 2.2cm 0cm 5.0cm, clip=true, height=2.5in, width=3.5in]{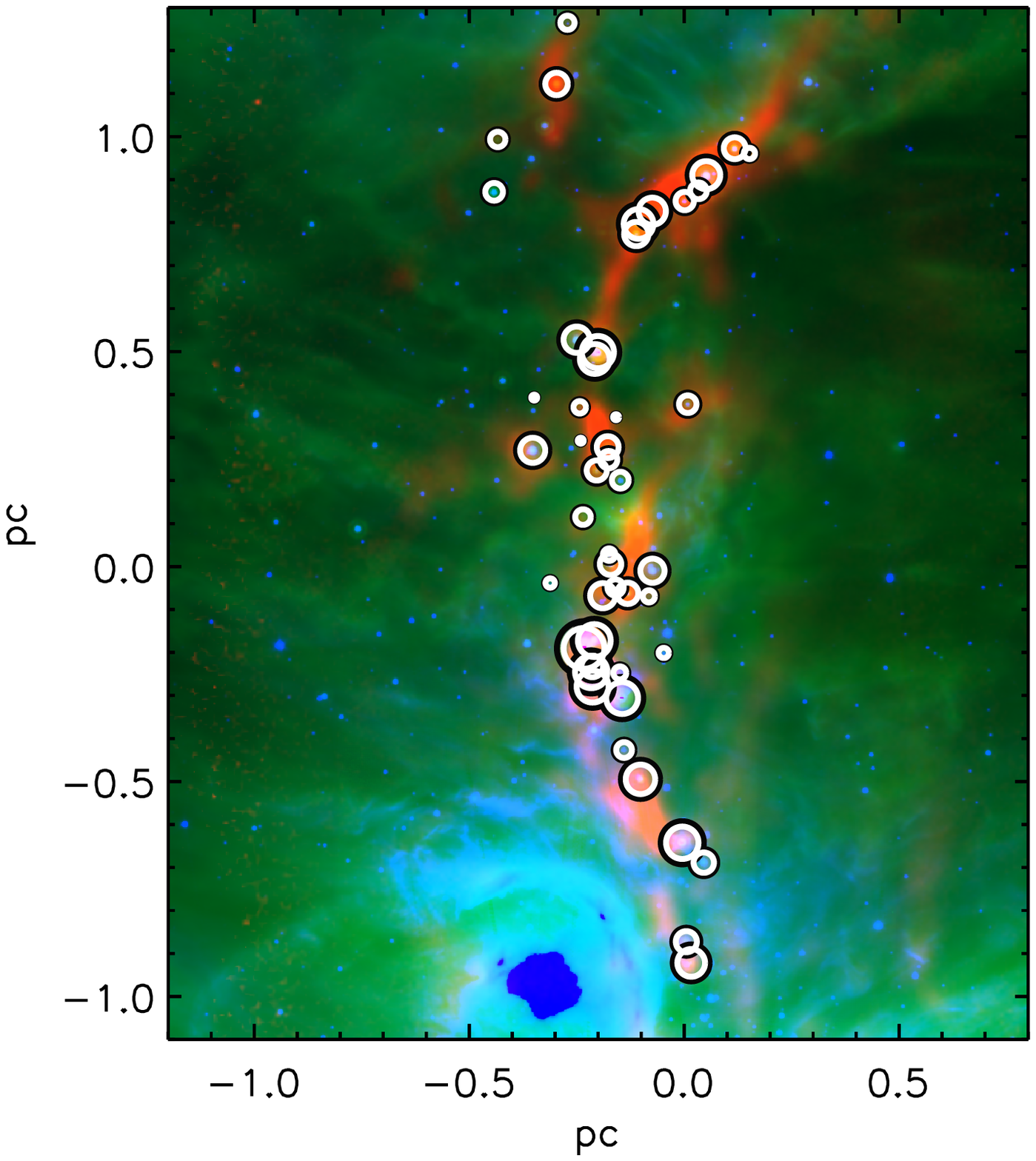}
\includegraphics[trim=0cm 0.7cm 0cm 5.0cm, clip=true, height=2.8in, width=3.5in]{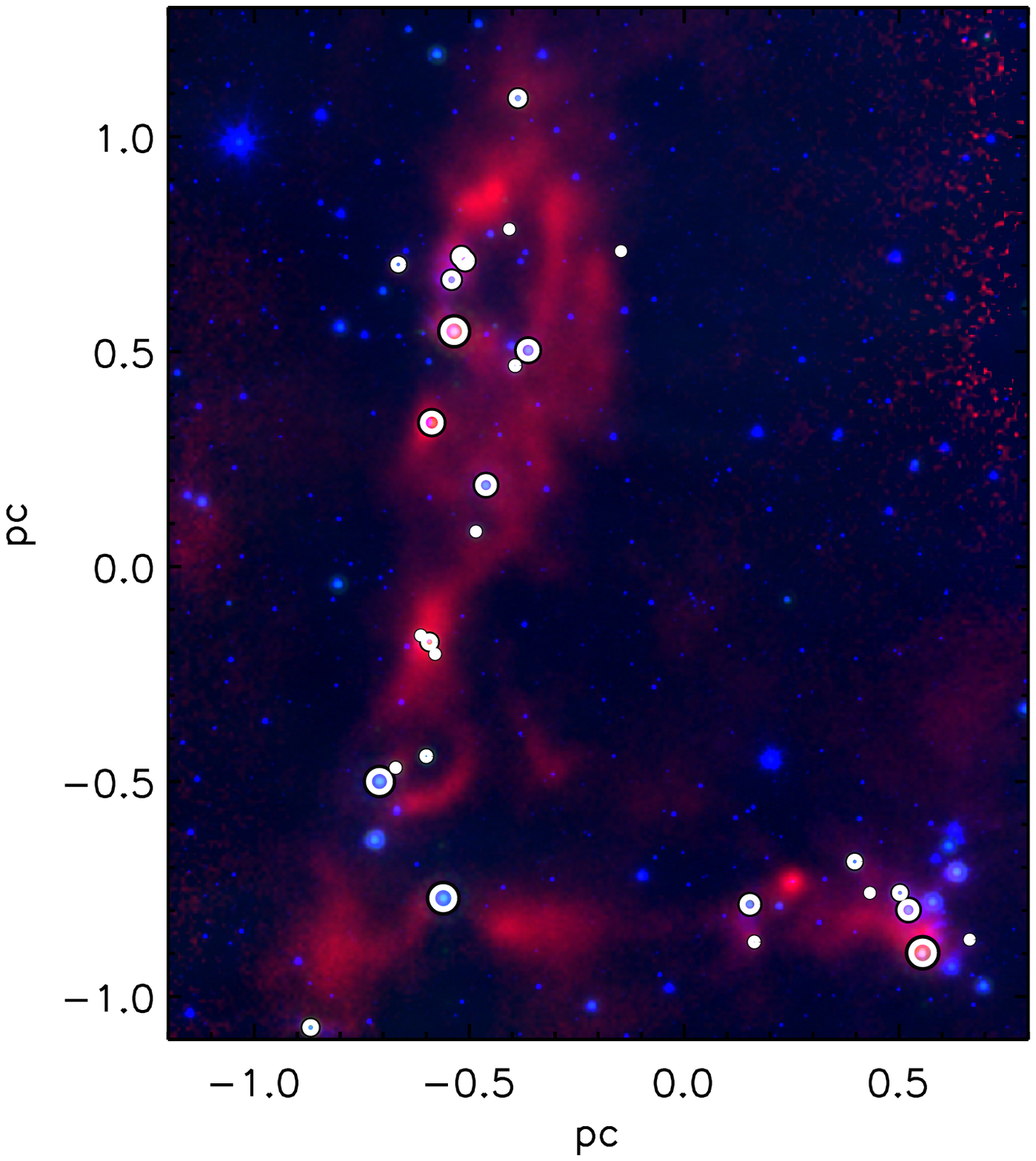}
\caption{\small Three-color images showing {\it Spitzer} 3.6 \um\ (blue), {\it Spitzer} 24 \um\ (green), and APEX/LABOCA 870 \um\ (red) of OMC-2/3 (top) and L1641S (bottom).  The APEX/LABOCA 870 \um\ will be presented by Stanke et al., in prep.   Overlaid are the position of the protostars; the size of the symbol is proportional to the log of the luminosity determined from the HOPS data.  }
\label{fig:hops_comp}
\end{figure}

\subsection{What Fraction of Protostars May be Interacting?}

With most protostars concentrated in regions of high gas column density and high stellar surface density, interactions between protostars may be common.  \citet{2007ApJ...669..493W} found that the protostars in the Serpens main cluster have a median projected separation of 8000 AU (corrected to the revised distance for Serpens of 429 pc \citep{2011RMxAC..40..231D}); smaller than the typical diameter of cores of 10,000-20,000 AU \citep{2008ApJ...684.1240E}.  They also found that the low relative velocities of the protostars and high gas density of the surrounding cloud core are consistent with the competitive accretion models of \citet{2006MNRAS.370..488B}, where the protostars compete for gas from a common reservoir. The projected spacing of the protostars in OMC2/3 is also smaller than the typical diameters of cores \citep[Fig.~\ref{fig:hops_comp}, also see][]{2013ApJ...763...57T,2013ApJ...768L...5L}. {\it Megeath et al.} (in prep.) examined the spacing between the {\it Spitzer} identified protostars in the Orion molecular clouds.  They found that  11\% and 23\% of the protostars are within a projected separation of  5000 and 10,000 AU from another protostar, respectively.  Thus, sources separated by $\le 10,000$~AU would potentially be able to interact through the collisions of their envelopes and tidal forces.  However, since the actual separations are larger than the measured projected separations, these percentages should be considered upper limits.  On the other hand, dynamical motions in such highly clustered regions \citep[e.g.,][]{2012MNRAS.427..637P} may increase the percentage of sources that interact.

\subsection{Do the Properties of Protostars Depend on their Environment?}

In their study of the protostellar luminosity distributions (hereafter: PLDs) in nine nearby molecular clouds, \citet{2012AJ....144...31K} searched for variations between the high and low density environments within a single cloud.  By dividing  protostars into two samples based on the local surface density of YSOs, they compared PLDs for protostars located in crowded, high density regions to those  in regions of low densities.  In the Orion molecular clouds, they found the PLDs of the high and low stellar density regions had a very low probability of being drawn from the same parent distribution, with the PLD for the protostars in denser environments biased to higher luminosities.  Thus, within the Orion molecular clouds, the luminosities of the protostars appears to depend on the surface density of the YSOs in the surrounding environment. Unfortunately, the smaller numbers of protostars found in the other eight molecular clouds precluded a definitive comparison of the high and low density regions in those clouds.  

One weakness of the Kryukova et al. study is the use of an empirical relationship to extrapolate from the infrared to bolometric luminosity.  However, using the full SEDs from  HOPS (Fig.~\ref{f.sed}) to directly determine the bolometric luminosity of the protostars in Fig.~\ref{fig:hops_comp}; the median $L_{bol}$ is 6.2~$L_{\odot}$  in the densely clustered OMC2/3 region and 0.7~$L_{\odot}$ in lower density L1641S region. This corroborates the result of \citet{2012AJ....144...31K} by showing the more densely clustered OMC2/3 region has systematically higher luminosity protostars.  Furthermore, it supports an interconnected relationship between the column density of the gas, the luminosities of the protostars, and the surface density of the protostars: filaments with higher gas column density will have both  a higher density of protostars and more luminous protostars.   In the near future, {\it Herschel} and ground-based (sub)millimeter imaging of protostellar environments should revolutionize our understanding of how the properties of protostars depend on their environment, and thereby place strong constraints on models of protostellar evolution.

\section{\textbf{SUMMARY}}


We have reported on recent progress in finding, identifying, and 
characterizing protostars in nearby clouds using infrared surveys
and follow-up studies. 
The coverage, sensitivity, and resolution of {\it Spitzer} and {\it Herschel}
have revealed objects of very low (\lint\ $< 0.1$ \lsun) luminosity
(VeLLOs), objects with very red SEDS (PBRS), and candidates for
first hydrostatic cores (FHSCs).
Based on the full census of YSOs found in the c2d, Gould Belt, Orion, and 
Taurus surveys, 
classified as in \citet{1994ApJ...434..614G}, and an assumed lifetime
for Class II objects of 2 Myr, we infer Class 0$+$I lifetimes of 0.42 to 
0.54 Myr, with the longer estimate applying to the Gould Belt clouds. 
While some differences in identification and classification remain,
we have attempted to minimize them. 

The luminosity distribution is very broad.
After reviewing models of evolution, focusing
on the accretion histories, we compared the luminosity distributions
from the models to the observations; either mass dependent or episodic
accretion models can match the data, but the isothermal sphere, constant
accretion rate, model cannot. Evidence of protostellar luminosity
variability, outflow episodicity, and ice evolution all support 
some form of episodic accretion. The remaining questions are how extreme
the variations can be, what the effect of variations are,
 and how much of the mass of the protostar is
accreted during episodes of high accretion.
Challenges for the future are to constrain short-term (decade scale) 
variability as a possible clue to longer term (e.g., $10^3$ yr) variability, 
and to ultimately determine the relative importance of stochastic (episodic) 
and secular processes in building the final masses of stars.

The first stage of star formation, the first hydrostatic core, remains
elusive. There are a number of candidates, but more theoretical work 
is needed to pin down the expected characteristics and to understand
the evolutionary status of the other novel objects, such
as PBRS and VeLLOs.

The existence and properties of disks in the protostellar phase are
beginning to yield to observational scrutiny, and ALMA will provide
a major advance. Current evidence suggests early formation of relatively
massive disks, although only a small number of detections have been 
made to date.  
Theories without magnetic fields predict more massive
disks than seen, while theories with rotation axes and magnetic fields aligned
struggle to create disks at all. Further theoretical
exploration is needed to understand the relationship of mass infall rate
from the core, its transition to the disk, and the processes that
allow it to accrete onto the growing star. Typical Stage II disks process
matter about 100 times more slowly than matter falls in during Stage I,
so faster processes are needed in Stage I to avoid mass build-up and 
instabilities.

Material falling onto the disk undergoes substantial chemical and, in some
cases, mineralogical evolution. Dust grains grow and may become crystallized,
while ice in mantles evolves chemically. Distillation of pure CO$_2$ ice
provides clues to the luminosity evolution. The state of ice and gas 
arriving at the disk set the stage for later chemical evolution during
Stage II.

The complete surveys reported here reveal the location of star formation 
within molecular clouds. Prestellar cores and protostars are highly 
concentrated in regions of high surface, and presumably volume, density
(see the accompanying chapter by {\it Padoan et al.}). The protostars in the 
very crowded environments of Orion appear to be systematically more luminous, 
and may be interacting in the regions of highest protostellar density.  
The ``clumps" hosting cluster formation are generally very filamentary, and
rich clusters prefer the nexus of filaments for their maternity wards.

The future is bright. ALMA will allow studies of unprecedented detail
of the density, temperature, and velocity fields in infalling envelopes 
\citep[e.g.,][]{2012A&A...544L...7P}.
Studies of embedded disks will clarify their masses and sizes, 
and detection
of Keplerian motions will finally constrain the masses of the growing
stars. JWST will allow deeper studies of the shorter (infrared) wavelenths
of deeply embedded objects, and SOFIA will provide spectroscopic
data on the brighter sources. On the theoretical side,
modelers should use physically realistic calculations to 
predict self-consistently all the observations:
the IMF and its realization in different regions, 
the structure, velocity field, and chemistry of infalling envelopes, 
the luminosity distribution, protostellar disk sizes and masses, 
the diversity of disks emerging in the Stage II phase,
variability measurements, protostellar masses (measured with 
Keplerian disks with ALMA and NOEMA, the successor to the PdBI), 
and environmental dependencies.

\noindent \textbf{ Acknowledgments.} 
The authors thank R.~Launhardt, J.~Green, 
and H.~G.~Arce for providing data for 
Fig.~\ref{f.sed} and Fig.~\ref{fig_hh46}, and the anonymous referee for 
comments that have improved this review.  
This review is based primarily on observations made with the 
{\it Spitzer Space Telescope} and {\it Herschel Space Observatory}.  
{\it Spitzer} is operated by the Jet Propulsion Laboratory, California 
Institute of Technology under a contract with NASA, and {\it Herschel} 
is an ESA space observatory with science instruments provided by European-led 
Principal Investigator consortia and with important participation from NASA.  
This review has made use of NASA's Astrophysics Data System Bibliographic 
Services.  
MMD, STM and WJF acknowledge support from NASA through awards issued by 
JPL/Caltech, and MMD acknowledges NSF support through grant AST-0845619 to 
Yale University.
The work of AMS was supported by the Deutsche Forschungsgemeinschaft
priority program 1573 (``Physics of the Interstellar Medium'').  
NJE acknowledges support from the NSF through grant AST-1109116 to the 
University of Texas at Austin.  
CAP acknowledges support provided by the NASA Astrobiology Institute through 
contract NNA09DA80A.  
JT acknowledges support provided by NASA through Hubble Fellowship 
grant \#HST-HF-51300.01-A awarded by the Space Telescope Science Institute, 
which is operated by the Association of Universities for Research in Astronomy, 
Inc., for NASA, under contract NAS 5-26555.  
EIV performed numerical simulations on the SHARCNET, ACEnet, VSC-2 scientific 
computer clusters, and acknowledges support from RFBR grant 13-02-00939.

\bibliographystyle{ppvi_lim1.bst}
\bibliography{ppvi_dunham}

\end{document}